\documentclass[11pt]{article}

\usepackage{tikz}

\usepackage{hyperref}
\usepackage[utf8]{inputenc}
\usepackage{color}
\usepackage{amsmath,amsfonts,amssymb,graphicx}
\usepackage{amsthm}
\usepackage[top=2.8cm,bottom=2.8cm,left=2.5cm,right=2.5cm]{geometry}
\usepackage{algorithm}
\usepackage{algorithmic}
\usepackage{soul}

\usepackage[toc]{appendix}
\usepackage{multirow}
\usepackage{enumitem}

\usepackage{authblk}
\usepackage{siunitx}
\usepackage{comment}
\usepackage{todonotes}

\newtheorem{remark}{Remark}

\newtheorem*{example*}{Example}

\newcommand{\bt}{\boldsymbol{\theta}}
\newcommand{\ba}{\boldsymbol{\alpha}}

\title{
Non-degenerate Marginal-Likelihood Calibration 
 \\with Application to Quantum Characterization} 

\author[1]{Mohammad Motamed \thanks{\texttt{motamed@unm.edu}}}
\affil[1]{{\small{Department of Mathematics and Statistics, University of New Mexico, Albuquerque, NM 87131}}}
\author[2]{N.~Anders Petersson \thanks{\texttt{petersson1@llnl.gov}}}
\affil[2]{{\small{Center for Applied Scientific Computing, Lawrence Livermore National Laboratory, Livermore, CA 94550}}}

\date{May 20, 2025}

\newcommand{\xb}{\boldsymbol{x}}

\newcommand{\rgl}{\rangle}

\begin{document}

\maketitle

\hrule
\bigskip

\begin{abstract}
We propose a marginal likelihood strategy within the Kennedy-O'Hagan (KOH) Bayesian framework, where a Gaussian process (GP) models the discrepancy between a physical system and its simulator. Our approach introduces a novel marginalized likelihood by integrating out the degenerate eigenspace of the covariance matrix, rather than approximating the original likelihood. Unlike approximation methods that compromise accuracy for computational efficiency, our method defines an exact likelihood---distinct from the original but preserving all relevant information. This formulation achieves computational efficiency and stability, even for large datasets where the covariance matrix nears degeneracy. 
Applied to the characterization of a superconducting quantum device at Lawrence Livermore National Laboratory, the approach enhances the predictive accuracy of the Lindblad master equations for modeling Ramsey measurement data by effectively quantifying uncertainties consistent with the quantum data.
\end{abstract}

\bigskip
\hrule


\section{Introduction}
\label{sec:intro}

Bayesian uncertainty analysis of complex physical systems, described by parametric simulation models, is a challenging task, particularly when accounting for both experimental uncertainty and model-form uncertainty. 
Experimental uncertainty typically arises from system variability or measurement errors, while model-form uncertainty stems from limitations or inadequacies of the simulation model. 
The intricate nature of these uncertainties, as well as the difficulty in clearly distinguishing between their different types and sources, complicates their identification and treatment. Addressing these challenges remains an active and complex area of research in Bayesian statistics and uncertainty quantification.

In a Bayesian framework, experimental uncertainty is typically addressed by incorporating additive random noise into the parametric simulation model. This approach defines the conditional distribution, or likelihood, of observing the data---represented for instance by a multivariate Gaussian distribution. This likelihood is then used in Bayes' formula to infer the unknown parameters of the simulation model. 
The conventional approach to account for model-form uncertainty involves correcting the simulation model using a discrepancy function, often modeled by a Gaussian process \cite{Kennedy_OHagan:2000,Kennedy_OHagan:01,Goldstein_Rougier:2004,Higdon_etal:2004,Bayarri_etal:2007,Higdon_etal:2008,Han_etal:2009,Brynjarsdottir_OHagan:2014,Ling_etal:2014,Sankararaman_Mahadevan:2015,Maupin_Swiler:2020}. This method, originally proposed by Kennedy and O'Hagan \cite{Kennedy_OHagan:2000,Kennedy_OHagan:01}, integrates into the Bayesian framework where experimental data is used to simultaneously estimate the parameters in the simulation model and the hyper-parameters in the discrepancy function, given prior distributions on both. 
This integrated Bayesian framework is commonly known as the KOH framework. 
Following the seminal work of Kennedy and O'Hagan, various Bayesian approaches have been proposed to incorporate model discrepancy. These approaches mainly differ in how they formulate the discrepancy---either as an ``external" term explicitly added on top of the simulation model \cite{Goldstein_Rougier:2004,Higdon_etal:2004,Bayarri_etal:2007,Higdon_etal:2008,Han_etal:2009,Brynjarsdottir_OHagan:2014,Ling_etal:2014,Sankararaman_Mahadevan:2015,Maupin_Swiler:2020} or as an ``internal" term embedded within the simulation model \cite{Oliver_etal:2015,Sargsyan_etal:2015,Portone_etal:2017,Morrison_etal:2018,Sargsyan_etal:2018}. 
The choice of approach and the treatment of model-form uncertainty generally depend on the specific problem at hand, with each strategy presenting its own set of challenges, limitations, and advantages. We note that, while this is an interesting research topic, a detailed discussion of these differences is beyond the scope of the present work.

This work makes two key contributions: (i) addressing a computational challenge in uncertainty quantification and (ii) applying the proposed methodology to quantum characterization. 
First, we address a computational challenge within the KOH framework, where a Gaussian process is employed as a prior to model the discrepancy between a physical system and its simulator. In moderate-to-large dataset regimes, the covariance matrix underlying the likelihood can approach degeneracy, rendering its evaluation computationally costly and numerically unstable. To address this challenge, we propose a stable and efficient sampling strategy that integrates out the degenerate eigenspace of the covariance matrix, yielding a novel marginalized likelihood. 
This approach yields an exact likelihood that, while distinct from the original, retains all relevant information, ensuring an effective marginal-likelihood KOH framework. By enhancing computational efficiency and stability, our method improves the feasibility and accuracy of simulation models for large-scale data. 
We note that while Bayesian analysis is often employed to address challenges arising from limited data availability, this work focuses on the moderate-to-large data regime, where abundant data introduces distinct computational difficulties, such as numerical instability, particularly under complex uncertainty structures. This is especially relevant in emerging fields like quantum characterization, where large datasets and correlated uncertainties predominate. Here, uncertainty quantification is driven not by limited data, but by intricate model-form and experimental uncertainties, making our approach relevant across a wide range of applications.

Second, we apply the proposed approach to characterize a tantalum-based transmon quantum device within the Quantum Device and Integration Testbed (QuDIT) system at Lawrence Livermore National Laboratory (LLNL) \cite{Place_etal:2021}. In quantum characterization, important physical properties of a quantum system, such as decoherence time scales and resonance frequencies, are estimated by a set of experimental protocols, including Ramsey experiments. 
In a Ramsey protocol, a sequence of pulses is first applied to the system, followed by a ``dark time" of a specified duration where no pulses are applied, and completed by a second sequence of pulses. The resulting state of the system is then measured, causing the quantum system to collapse into one of its classical states. The protocol is repeated many times, and the probability of observing a specific state is estimated by its frequency.

We formulate the quantum characterization problem within the KOH Bayesian framework, deriving a multivariate Gaussian likelihood function based on the parameters of the quantum dynamical model and the hyperparameters of the Gaussian process. 
Given the challenges posed by the large data-set regime, where the likelihood function for quantum system measurements becomes near-degenerate, we utilize a marginalized likelihood technique to infer parameters in the quantum dynamical model. 
Our results demonstrate how incorporating model discrepancy enhances the predictive capability of quantum simulation models, specifically Lindblad's master equations \cite{Lindblad:1976}, using quantum measurements. 
This approach, enabled by the marginalized likelihood within the KOH framework, is particularly significant as most existing Bayesian techniques for quantum characterization (e.g., \cite{Wiebe_etal:2015,Wang_etal:2017,Ralph_etal:2017,Evans_etal:2019,Bennink_etal:2019,Gentile_etal:2021,Peng_etal:2023}) only account for experimental uncertainty, assuming that quantum dynamical models accurately reflect the system's dynamics. 
To the best of our knowledge, existing approaches in quantum characterization neglect model-form uncertainty, even though it can significantly impact predictive uncertainty and overall model fidelity.

The remainder of the paper is organized as follows. Section \ref{sec:general} reviews the KOH Gaussian process-based Bayesian framework in a general setting. Section \ref{sec:marginal_likelihood} introduces the proposed marginalized likelihood approach for large datasets, along with a theoretical validation and a strategy for stable and efficient Markov chain Monte Carlo sampling. This section also includes a discussion of related work and highlights the novelty of our approach. In Section \ref{sec:quantum_char}, we apply our method to quantum characterization using measurement data from the QuDIT device. Concluding remarks are presented in Section \ref{sec:conclusion}.

\section{An overview of the KOH Bayesian approach}
\label{sec:general}

Assuming ample amounts of relevant experimental observations, a general approach for uncertainty analysis and predictive computing of complex physical systems, described by parametric simulation models, 
involves two key steps. First, physical observations are used to infer the unknown model parameters within a Bayesian framework, a process known as model calibration or characterization. Next, the calibrated model, incorporating the inferred uncertain parameters, predicts the unobserved behavior of the physical system. The KOH Bayesian approach accomplishes this by utilizing Gaussian variables to represent experimental noise and a Gaussian process to model the discrepancy between the physical system and the simulation. The calibrated model, along with the adjusted discrepancy function, is then used for making predictions. In this section, we review the two key steps of the KOH Bayesian approach as applied to a dynamical system.

Consider a set of noisy experimental measurements $\{ y_1, \dotsc, y_n \}$ corresponding to a set of time points $\{ t_1, \dotsc, t_n \}$, with $0 \le t_1 < t_2 < \dotsc < t_n \le T$. 
Let $f(t_i; \bt)$ represent the simulated measurement outcome corresponding to time $t_i$, obtained by evaluating a simulation model that describes the system's dynamics, with parameters $\bt \in \Theta \subset {\mathbb R}^d$. For example, in the quantum application considered here, each measurement $y_i$ represents the probability (or population) of measuring the quantum system in one of its quantized states, such as ${|0\rangle, |1\rangle, |2\rangle }$, corresponding to the ``dark time'' $t_i \in [0,T]$. 
In this context, $f(t_i; \bt)$ could represent the simulated population obtained by solving Lindblad's master equations with the parameters $\bt$, including decoherence time scales and resonance frequencies. 
Thus, we are facing an ``inverse" problem of recovering $\bt$ from a set of noisy observations ${\bf y}_n := \{ y_i \}_{i=1}^n$, based on the model ${\bf f}(\bt) := \{ f(t_i; \bt) \}_{i=1}^n$. 
For more details on this particular application, see Section \ref{sec:quantum_char}.

In a Bayesian framework, the parameters $\boldsymbol\theta$ are treated as random variables characterized by a joint probability distribution, known only through a set of observations ${\bf y}_n$ corrupted by noise (see, e.g., \cite{Gelman_etal:04}). By positing a likelihood function $\Pr({\bf y}_n | \bt)$, which represents the conditional distribution of the observed data given $\bt$ through the model ${\bf f}(\bt)$, and a prior distribution $\Pr(\bt)$ that reflects our belief about $\boldsymbol\theta$ before observing the data, we can determine the conditional posterior distribution of $\bt$ using Bayes' rule \cite{Bayes:1763}:
$$
\Pr(\bt | {\bf y}_n) = \frac{\Pr({\bf y}_n | \bt) \, \Pr(\bt)}{\int_{\Theta}  \Pr({\bf y}_n | \bt) \, \Pr(\bt) \, d \bt}.
$$

In the KOH approach \cite{Kennedy_OHagan:01}, the relationship between the experimental and simulated measurements is modeled as: 
\begin{equation}\label{data_model}
y_i = f(t_i; \bt) + \delta(t_i) + \varepsilon_i, \qquad i=1, \dotsc, n,
\end{equation}
where $\varepsilon_i$ represents the random noise in the $i$-th observation $y_i$, and $\delta(\cdot)$ is the model discrepancy function. Following common practice in Bayesian analysis, we assume that the $\varepsilon_i$'s are independently and identically distributed Gaussian variables with mean zero and variance $\sigma_{\varepsilon}^2$:
$$
\varepsilon_i \sim \mathcal{N}(0,\sigma_{\varepsilon}^2), \qquad i=1, \dotsc, n.
$$
The discrepancy function $\delta(\cdot)$ is modeled by a Gaussian process, which is an infinite collection of random variables where any finite subset has a joint Gaussian distribution characterized by a mean and covariance, both determined by the mean and covariance kernel of the process (see, e.g., \cite{Rasmussen_Williams:06}). While various forms of covariance kernels can be employed, here we consider a Gaussian process with zero mean and an exponential covariance kernel:
\begin{equation}\label{specific_covariance}
\delta(t) \sim {\mathcal {GP}}(0, \kappa(t,t')), 
\qquad 
\kappa(t,t') = \sigma_{\delta}^2 \, \exp \left( -\frac{|t - t'|^{\gamma}}{2 \, \tau^{\gamma}} \right), 
\end{equation}
where $\sigma_{\delta}^2$ is the variance, $\tau$ is the time scale, and $\gamma > 0$ is a fixed exponent. In the numerical experiments presented in Section \ref{sec:quantum_char}, we use a linear exponential covariance kernel with $\gamma = 1$. 
We denote by $K$ the $n \times n$ matrix of the covariance kernel $\kappa$ evaluated at all pairs of (training) time points:
$$
K = [K_{i,j}] = [\kappa(t_i,t_j)] \in {\mathbb R}^{n \times n}, \qquad i,j \in \{ 1, \dotsc, n \}.
$$
The covariance kernel considered in \eqref{specific_covariance} is commonly used in scenarios where the correlation between data points diminishes as their temporal separation increases. However, it is important to note that our approach in Section \ref{sec:marginal_likelihood} is applicable to general covariance kernels and is not limited to the specific form given in \eqref{specific_covariance}.

From the assumption that the difference between experimental and simulated data equals the sum of model discrepancy and experimental noise (see \eqref{data_model}), and the definition of a Gaussian process, it follows that the data points $\{y_1, \dotsc, y_n \}$ can be viewed as the coordinates of a Gaussian random vector:
$$
{\bf y}_n = (y_1, \dotsc, y_n)\sim {\mathcal N}({\bf f}(\bt), \Sigma(\ba)), 
$$
with the mean vector ${\bf f}(\bt) \in {\mathbb R}^n$ and the covariance matrix $\Sigma(\ba) \in {\mathbb R}^{n \times n}$ given by:
$$
{\bf f}(\bt) = (f(t_1; \bt), \dotsc, f(t_n; \bt)), 
\qquad 
\Sigma(\ba) = K(\sigma_{\delta},\tau) + \sigma_{\varepsilon}^2 \, I_n. 
$$
Here, the hyper-parameter vector $\ba=(1/\sigma_{\varepsilon}^2, 1/\sigma_{\delta}^2, \tau)$ collects the reciprocal of the Gaussian noise variance $1/\sigma_{\varepsilon}^2$, the reciprocal of the Gaussian process variance $1/\sigma_{\delta}^2$, and the time scale $\tau$ of the Gaussian process. 
The likelihood function is then given by:
\begin{equation}\label{likelihood}
\Pr ({\bf y}_n | \bt, \ba) = (2 \pi)^{-n/2} \, |\Sigma(\ba)|^{-1/2} \exp \left( - \frac{1}{2} ({\bf y}_n - {\bf f}(\bt))^{\top} \Sigma^{-1}(\ba) \, ({\bf y}_n - {\bf f}(\bt)) \right). 
\end{equation}
Following Bayes' rule, the posterior distribution of the parameters given the data is:
\begin{equation}\label{Bayes_rule}
\Pr(\bt, \ba| {\bf y}_n) \propto \Pr({\bf y}_n | \bt, \ba) \, \Pr(\bt, \ba).
\end{equation}

Once the posterior distribution is expressed as in \eqref{Bayes_rule}, we can, in principle, generate samples from it using well-established computational tools, such as Markov Chain Monte Carlo (MCMC) methods; see, e.g., \cite{MCMC:04,Kaipo_Somersalo:05}. These methods enable us to approximate the posterior distribution and to perform subsequent inference and uncertainty quantification based on the generated samples.

\section{A marginal-likelihood KOH approach for large data sets}
\label{sec:marginal_likelihood}

In the case of large data sets ($n \gg 1$), two issues may arise when computing the likelihood function \eqref{likelihood}. First, evaluating the inverse and determinant of $\Sigma(\ba) \in \mathbb{R}^{n \times n}$ becomes computationally prohibitive, as this cost grows cubically with $n$. More critically, as the time points get closer to each other, the covariance matrix $\Sigma(\ba)$ becomes increasingly degenerate, causing its determinant to become exponentially small in $n$, leading to numerical round-off errors. 
To address these problems, we propose a stable approach that leverages the eigen-decomposition of the covariance matrix $\Sigma(\ba)$. This decomposition allows us to split its range (or column space) into two subspaces: one spanned by eigenvectors corresponding to large eigenvalues, and the other by eigenvectors corresponding to small eigenvalues. We then project all random variables orthogonally onto these two subspaces, followed by marginalizing out the random variables associated with the subspace corresponding to the small eigenvalues. 
This approach is robust and independent of the eigen-expansion of the covariance kernel, making it applicable to covariance kernels of any form, including the specific kernels in \eqref{specific_covariance} with any fixed exponent $\gamma > 0$. 
We also present an MCMC sampling algorithm tailored to this framework, along with a comparison to related methods and a discussion highlighting the novelty of our approach.

\subsection{A stable marginalized likelihood framework}
\label{sec:marginal}

We first note that in a sampling framework, such as MCMC, numerous evaluations of the likelihood function \eqref{likelihood} are required for various values of the parameter vector $(\bt,\ba)$. 
For ease of exposition, given fixed values of the parameters $(\bt,\ba)$, we define ${\bf x} := {\bf y}_n - {\bf f}(\bt)$ and $\Sigma := \Sigma(\ba)$. The likelihood function \eqref{likelihood} can then be expressed in condensed form:
\begin{equation}\label{eq:eigen_decomp}
L({\bf x}) = \frac{1}{C} \exp\left( -\frac{1}{2}{\bf x}^{\top} \Sigma^{-1} {\bf x} \right),\qquad 
C = \int_{\mathbb{R}^n} \exp\left( -\frac{1}{2}{\bf x}^{\top} \Sigma^{-1} {\bf x} \right)\, d{\bf x}.
\end{equation}
As mentioned earlier, we consider scenarios where the covariance matrix $\Sigma \in \mathbb{R}^{n\times n}$ is nearly degenerate, with many small eigenvalues. Direct computation of its determinant and inverse may thus lead to numerical instabilities and become impractical. However, since the covariance matrix is symmetric and positive definite, it possesses real and positive eigenvalues and a complete set of orthonormal eigenvectors. Consequently, the covariance matrix can be expressed through the following eigen-decomposition:
\begin{align*}
    \Sigma = E \Lambda E^{\top}, 
    \qquad E^{\top} E = I_n, 
    \qquad
    \Lambda = \begin{bmatrix}
       \Lambda_I & 0 \\ 0 & \Lambda_{II}
    \end{bmatrix}, \qquad 
    E = \begin{bmatrix}
        E_I & E_{II}
    \end{bmatrix}, 
\end{align*}
where the diagonal matrices $\Lambda_I = \mbox{diag}(\lambda_1,\ldots,\lambda_r)$ and $\Lambda_{II} = \mbox{diag}(\lambda_{r+1}, \ldots,\lambda_n)$ contain the eigenvalues arranged in descending order, and such that all eigenvalues in $\Lambda_{II}$ are small: $\max_{j\geq r+1} (\lambda_j) \leq \epsilon \ll 1$. The matrices $E_I \in \mathbb{R}^{n\times r}$ and $E_{II} \in \mathbb{R}^{n\times (n-r)}$ hold the corresponding eigenvectors associated with the eigenvalues in $\Lambda_I$ and $\Lambda_{II}$, respectively. Further, the orthogonality of the eigenvectors imply $E_I^{\top} E_{II} = {\bf 0} \in {\mathbb R}^{r \times (n-r)}$.

Next, to evaluate $L({\bf x})$, we need to solve the linear system $\Sigma \, {\bf z} = {\bf x}$. To do this efficiently, we introduce the orthogonal projection matrices:
\begin{align*}
    P_I = E_I E_I^{\top}, \qquad P_{II} = E_{II} E_{II}^{\top},\qquad P_I + P_{II} = I_n,
\end{align*}
and decompose the vectors ${\bf x} = P_I {\bf x} + P_{II} {\bf x}$ and ${\bf z} = P_I {\bf z} + P_{II} {\bf z}$ as follows:
\begin{align*}
    {\bf x} &= E_I {\bf x}_I + E_{II} {\bf x}_{II}, \qquad {\bf x}_I = E_I^{\top} {\bf x} \in \mathbb{R}^r, \qquad {\bf x}_{II} = E_{II}^{\top} {\bf x} \in \mathbb{R}^{n-r},\\
    {\bf z} &= E_I {\bf z}_I + E_{II} {\bf z}_{II}, \qquad {\bf z}_I = E_I^{\top} {\bf z} \in \mathbb{R}^r, \qquad 
    {\bf z}_{II} = E_{II}^{\top} {\bf z} \in \mathbb{R}^{n-r}.
\end{align*}
It is straightforward to see that the linear system $\Sigma {\bf z} = {\bf x}$, or equivalently $\Lambda E^{\top} {\bf z} = E^{\top} {\bf x}$, decomposes into two smaller linear systems with diagonal matrices:
\begin{align*}
    \Lambda_I {\bf z}_I = {\bf x}_I,\qquad \Lambda_{II} {\bf z}_{II} = {\bf x}_{II}.
\end{align*}
Therefore, the solution of the system $\Sigma \, {\bf z} = {\bf x}$ can be written as:
$$
{\bf z} = E_I {\bf z}_I + E_{II} {\bf z}_{II}, \qquad {\bf z}_I = \Lambda_I^{-1} E_I^{\top} {\bf x}, \qquad {\bf z}_{II}= \Lambda_{II}^{-1} E_{II}^{\top} {\bf x}.
$$
The likelihood function can be decomposed according to
\begin{align*}
L({\bf x}) = \frac{1}{C} \exp \left( -\frac{1}{2} {\bf x}^{\top} \Sigma^{-1} {\bf x} \right) 
& = \frac{1}{C} \exp \left( -\frac{1}{2} {\bf x}^{\top} {\bf z} \right)  \\
& = \frac{1}{C} \exp \left( -\frac{1}{2} {\bf x}^{\top} ( E_I \Lambda_I^{-1} E_I^{\top} {\bf x} + E_{II} \Lambda_{II}^{-1} E_{II}^{\top} {\bf x}) \right) \\
& = \underbrace{\frac{1}{C_I} \exp \left( -\frac{1}{2} {\bf x}_I^{\top} \Lambda_I^{-1} {\bf x}_I \right)}_{L_I ({\bf x}_I)} \ 
\underbrace{\frac{1}{C_{II}} \exp \left( -\frac{1}{2} {\bf x}_{II}^{\top} \Lambda_{II}^{-1} {\bf x}_{II} \right)}_{L_{II}(\xb_{II})},
\end{align*}
where $C_I = (2\pi)^{r/2} \, |\Lambda_I |^{1/2}$ and $C_{II} = (2\pi)^{(n-r)/2} \, | \Lambda_{II} |^{1/2}$. 
In other words, we have decomposed the Gaussian function $L$ into the product of two Gaussian density functions $L_I$ and $L_{II}$. 
Notably, while the computation of $L_I({\bf x}_I)$ with ${\bf x}_I = E_I^{\top} {\bf x}$ is stable and efficient (especially when $r \ll n$), the computation of $L_{II}({\bf x}_{II})$ with ${\bf x}_{II} = E_{II}^{\top} {\bf x}$ may suffer from numerical instabilities. 
This instability arises because the diagonal elements of $\Lambda_{II}$ are very small, making the computations of its inverse and determinant prone to round-off errors. To circumvent these numerical instabilities, we marginalize the joint likelihood function $L({\bf x}) = L_I({\bf x}_I) \, L_{II}({\bf x}_{II})$ over ${\bf x}_{II}\in \mathbb{R}^{n-r}$, which results in the marginal likelihood:
$$
\int_{\mathbb{R}^{n-r}} L({\bf x}) \, d{\bf x}_{II} = 
\int_{\mathbb{R}^{n-r}} 
L_I({\bf x}_I) \, L_{II}({\bf x}_{II}) \, d{\bf x}_{II} =
L_I({\bf x}_I) \, \int_{\mathbb{R}^{n-r}} 
 L_{II}({\bf x}_{II}) \, d{\bf x}_{II} = L_I({\bf x}_I).
$$

The above procedure amounts to replacing \eqref{likelihood} with the following marginal likelihood: 
\begin{equation}\label{marginal_likelihood}
\Pr({\bf y}_n | \bt, \ba) = 
(2 \pi)^{-r/2} \, |\Lambda_I(\ba)|^{-1/2} \exp \left( - \frac{1}{2} \left( E_I^{\top}(\ba) ({\bf y}_n - {\bf f}(\bt)) \right)^{\top} \Lambda_I^{-1}(\ba) \, \left( E_I^{\top}(\ba) ({\bf y}_n - {\bf f}(\bt)) \right) \right),
\end{equation}
where $\Lambda_I(\ba)$ and $E_I(\ba)$ contain the $r$ largest eigenvalues and corresponding eigenvectors of the covariance matrix $\Sigma(\ba)$. 
When $r\ll n$, since $\Lambda_I$ is of small size and does not have any small eigenvalues, the marginal likelihood \eqref{marginal_likelihood} can be efficiently evaluated without numerical round-off problems, even when the covariance matrix $\Sigma$ is nearly singular.

\begin{remark}We emphasize that the marginal likelihood $L_I({\bf x}_I)$ should not be interpreted as an approximation of the original likelihood $L({\bf x}) = L_I({\bf x}_I) \, L_{II}({\bf x}_{II})$. While $L_{II}({\bf x}_{II})$ has unit integral, it is not necessarily close to 1. In this regard, our approach is distinct from techniques that attempt to approximate either the original likelihood model, or the inverse of the covariance function. 
Our approach is fundamentally information-driven, ensuring that the essential statistical content of the data is preserved, because the crucial information resides in $L_I({\bf x}_I)$. 
For further details, see Section \ref{subsec:preservation}. 
\end{remark}

\paragraph{Dominant eigenvalue decomposition.}
The proposed strategy requires computing the $r$ largest eigenvalues and their corresponding eigenvectors of the covariance matrix. 
Randomized methods for eigenvalue decomposition, such as randomized singular value decomposition, randomized power iteration, and randomized block Krylov subspace methods, offer a computationally efficient approach for large-scale matrices (see, e.g., \cite{Halko_etal:2011, Mahoney:2011, Martinsson_Tropp:2020} for a review). {We note that the strategy remains efficient when $r$ is small relative to $n$, which is common when the data can be well-represented by a few principal components.}

\subsection{Preservation of Statistical Information via Marginalization}
\label{subsec:preservation}

This subsection elucidates why the marginalized likelihood retains the essential statistical information of the original GP model within the KOH framework. Unlike approximation methods that trade accuracy for computational efficiency, our approach marginalizes over the degenerate eigenspace of the covariance matrix, yielding an exact likelihood over a reduced, non-degenerate subspace. This preserves the model's inferential and predictive capabilities while enhancing computational stability, particularly when the covariance matrix approaches singularity.

Following the notation in Section~\ref{sec:marginal}, we consider a GP prior model where the observations \( \mathbf{x} := \mathbf{y}_n - \mathbf{f}(\bt)  \in \mathbb{R}^n \) follow \( \mathbf{x} \sim \mathcal{N}(\mathbf{0}, \Sigma) \), with $\mathbf{x}$ representing the measured data \( \mathbf{y}_n \) and mean function \( \mathbf{f}(\bt) \). We focus on cases where the covariance matrix \( \Sigma \in \mathbb{R}^{n \times n} \) has rank  $r < n$ due to zero or nearly zero eigenvalues (i.e., $\epsilon \approx 0$, or $\Lambda_{II}$ being effectively a zero matrix). Instead of approximating the original model, our approach integrates out this degenerate eigenspace, preserving the inferential power of the GP while ensuring numerical stability.

\paragraph{Sufficiency of the Non-Degenerate Subspace.} 
Given \( \Sigma \)’s rank \( r < n \), its inverse does not exist in the classical sense, necessitating the Moore-Penrose pseudoinverse \( \Sigma^+ \), computed via the eigendecomposition (cf. Section~\ref{sec:marginal}):
\[
\Sigma = E \Lambda E^\top, \quad \Sigma^+ = E \Lambda^+ E^\top,
\]
where \( E = [E_I \, E_{II}] \), \( \Lambda = \text{diag}(\Lambda_I, \Lambda_{II}) \), \( \Lambda_I = \text{diag}(\lambda_1, \ldots, \lambda_r) \) contains the \( r \) positive eigenvalues, and \( \Lambda_{II} = 0 \) is the \( (n-r) \times (n-r) \) zero matrix. Thus, \( \Lambda^{+} = \text{diag}(\Lambda_I^{-1}, 0) \), with \( (\Lambda_I^{-1})_{ii} = 1/\lambda_i \) for \( i = 1, \ldots, r \).

Since the eigendecomposition of $\Sigma$ reduces to \( \Sigma = E_I \Lambda_I E_I^{\top} \), the covariance is fully determined by the non-degenerate subspace spanned by \( E_I \). The degenerate subspace, spanned by \( E_{II} \), has zero variance, as \( \Lambda_{II} = 0 \). Projecting the data onto these subspaces, \( {\bf x}_I = E_I^{\top} {\bf x} \) captures all variability encoded by \( \Sigma \), while \( {\bf x}_{II} = E_{II}^{\top} {\bf x} \) lies in the null space, where the GP prior assigns no variability. Hence, the statistical information relevant to the GP model resides entirely in \( {\bf x}_I \).

\paragraph{Connection to Likelihood Marginalization.} 
With zero variance in the null space (\( E_{II} \)), the original likelihood, 
\[
\Pr({\bf x}) = (2\pi)^{-n/2} |\Sigma|^{-1/2} \exp\left(-\frac{1}{2} {\bf x}^{\top} \Sigma^{-1} {\bf x}\right),
\]
is undefined since \( |\Sigma| = 0 \). Using \( \Sigma^{+} \) instead, the quadratic form becomes:
\[
{\bf x}^{\top} \Sigma^{+} {\bf x} = {\bf x}^{\top} E_I \Lambda_I^{-1} E_I^{\top} {\bf x} = {\bf x}_I^{\top} \Lambda_I^{-1} {\bf x}_I,
\]
indicating that the GP's likelihood, and thus its inference/prediction, is determined solely by \( \mathbf{x}_I \). Marginalizing over \( {\bf x}_{II} \), we obtain:
\[
\Pr({\bf x}_I) = (2\pi)^{-r/2} |\Lambda_I|^{-1/2} \exp\left(-\frac{1}{2} {\bf x}_I^{\top} \Lambda_I^{-1} {\bf x}_I\right).
\]
This is a new exact likelihood in the \( r \)-dimensional non-degenerate subspace. Since \( {\bf x}_{II} \) carries no prior uncertainty, integrating it out preserves the GP's predictive and inferential power within \( \Sigma \)'s range. Discarding \( {\bf x}_{II} \), orthogonal to this range, incurs no loss of information if the data-generating process lacks variation in those directions, as assumed in structurally degenerate cases. If variation exists outside this range, it lies beyond the degenerate GP's scope, requiring a different non-degenerate model.

\begin{remark}
One may ask what happens when the eigenvalues of \(\Lambda_{II}\) are small but strictly positive. In such cases, we emphasize that the reduced likelihood \(\Pr({\bf x}_I)\) should not be interpreted as an approximation of the full likelihood \(\Pr({\bf x})\), but rather as an exact marginal likelihood that captures the relevant statistical content of the degenerate GP model. The framework here is best understood through an information-theoretic lens: when the variance along certain directions is very small (i.e., eigenvalues \(\ll 1\)), the associated entropy is also small, indicating that the posterior distribution is sharply concentrated in a lower-dimensional subspace. While the marginalization is exact in the limiting case \(\Lambda_{II} = 0\), the information loss for \(\Lambda_{II} \approx 0\) is also negligible in practice, as the data components along these nearly degenerate directions contribute little uncertainty or predictive utility. A precise quantification of this loss remains an interesting direction for future work.
\end{remark}

\subsection{A Markov Chain Monte Carlo sampling algorithm}\label{sec:MHG_algorithm}

We now present a Metropolis-within-Gibbs sampling strategy to generate samples from the parameter posterior distribution $\Pr (\bt, \ba | {\bf y}_n)$, as described in \eqref{Bayes_rule}. This algorithm iteratively generates a sequence of samples to form a Markov chain, whose distribution converges to the target posterior distribution in the limit; for details on MCMC sampling techniques, see \cite{Gelman_etal:04} and related references. 
The sampling strategy consists of two interactive components: a Gibbs sampler and a Metropolis sampler. In each iteration, Gibbs sampling is used to sample from the joint distribution $\Pr (\bt, \ba | {\bf y}_n)$. 
Specifically, the algorithm first samples $\ba$ from $\Pr(\ba | \bt, {\bf y}_n)$ and then samples $\bt$ from $\Pr(\bt | \ba, {\bf y}_n)$, utilizing separate Metropolis steps for each parameter set. 
The detailed steps of the algorithm are summarized in Algorithm \ref{ALG_MG}.
\begin{algorithm}[!ht]
\caption{{\fontsize{11}{12}\selectfont
    Metropolis-within-Gibbs Sampling}}
\label{ALG_MG}
\begin{algorithmic} 
\medskip

\STATE {\bf 1.} {\it Input}: 

\ \ \ \ \ ${\bf y}_n:$  vector of experimental measurements. 

\ \ \ \ \ ${\bf f}(\bt):$  vector of simulated measurements. 

\ \ \ \ \ $\Sigma(\ba):$  covariance matrix.

\ \ \ \ \ $\Pr(\bt,\ba):$  prior on parameters $(\bt,\ba)$. 

\ \ \ \ \ $q_{\bt}, q_{\ba}:$  symmetric proposal distributions for $\bt$ and $\ba$.  

\ \ \ \ \ $M:$  number of samples. 

\ \ \ \ \ $r:$  number of dominant eigenvalues of the covariance matrix to retain.

\medskip
\STATE {\bf 2.} {\it Initialization}: 

\ \ \ \ \ Draw $(\bt^{(0)}, \, \ba^{(0)}) \sim \Pr(\bt,\ba)$.

\ \ \ \ \ Compute dominant eigenpairs $(\Lambda_I(\ba^{(0)}), E_I(\ba^{(0)}))$ of $\Sigma(\ba^{(0)})$.

\ \ \ \ \ Compute the likelihood $\Pr({\bf y}_n | \bt^{(0)}, \ba^{(0)})$ by \eqref{marginal_likelihood}.

\medskip
\STATE {\bf 3.} {\it Gibbs sampling loop}:

\medskip
\ \ \ \ {\bf for} $m=0, 1, \ldots, M -1$

\begin{itemize}[leftmargin=1.2cm]
\setlength\itemsep{-0.3em}
\item Metropolis sampling step 1: draw a sample $\ba^{(m+1)} \sim \Pr(\ba | \bt^{(m)}, {\bf y}_n)$ as follows:

\begin{itemize}[leftmargin=.4cm]
\setlength\itemsep{-0.3em}
\item[$\circ$] Generate a candidate sample $\tilde{\ba}$ drawn from proposal $q_{\ba}(\ba^{(m)},\tilde{\ba})$. 

\item [$\circ$] Compute dominant eigenpairs $(\Lambda_I(\tilde{\ba}), E_I(\tilde{\ba}))$ of $\Sigma(\tilde{\ba})$ and 
$\Pr({\bf y}_n | \bt^{(m)}, \tilde{\ba})$ by \eqref{marginal_likelihood}.

\item[$\circ$] Compute the acceptance probability 
\vspace{-0.2cm}
$$
\vspace{-0.2cm}
\gamma_{\ba} = \min \Bigg\{ 1, \ 
\frac{\Pr(\tilde{\ba} | \bt^{(m)}, {\bf y}_n)}{\Pr(\ba^{(m)}
  | \bt^{(m)}, {\bf y}_n)} 
  \Bigg\}
= 
\min \Bigg\{ 1, \ 
\frac{\Pr({\bf y}_n | \bt^{(m)},\tilde{\ba})}{\Pr({\bf y}_n | \bt^{(m)}, \ba^{(m)})} 
\ \cdot \ 
\frac{\Pr(\bt^{(m)},\tilde{\ba})}{\Pr(\bt^{(m)},\ba^{(m)})}
\Bigg\}.
$$

\item[$\circ$] Set \ \ 
$
\ba^{(m+1)} = \left\{ \begin{array}{l l}
\tilde{\ba} & \qquad \text{if} \ \  \gamma_{\ba} \ge u \sim \text{Uniform}(0,1), \\
\ba^{(m)} &  \qquad \text{otherwise}.
\end{array} \right.
$

\item[$\circ$] Update $(\Lambda_I(\ba^{(m+1)}), E_I(\ba^{(m+1)}))$ accordingly.
\end{itemize}

\item Metropolis sampling step 2: draw a sample $\bt^{(m+1)} \sim \Pr(\bt | \ba^{(m+1)}, {\bf y}_n)$ as follows:

\begin{itemize}[leftmargin=.4cm]
\setlength\itemsep{-0.3em}
\item[$\circ$] Generate a candidate sample $\tilde{\bt}$ drawn from proposal $q_{\bt}(\bt^{(m)},\tilde{\bt})$. 

\item [$\circ$] Compute 
$\Pr({\bf y}_n | \tilde{\bt}, \ba^{(m+1)})$ by \eqref{marginal_likelihood}.

\item[$\circ$] Compute the acceptance probability 
\vspace{-0.2cm}
$$
\vspace{-0.2cm}
\gamma_{\bt} = \min \Bigg\{ 1, \ 
\frac{\Pr(\tilde{\bt} | \ba^{(m+1)}, {\bf y}_n)}{\Pr(\bt^{(m)}
  | \ba^{(m+1)}, {\bf y}_n)} 
  \Bigg\}
= 
\min \Bigg\{ 1, \ 
\frac
{\Pr({\bf y}_n | \tilde{\bt}, \ba^{(m+1)})}{\Pr({\bf y}_n |\bt^{(m)},\ba^{(m+1)})} 
\ \cdot \ 
\frac{\Pr(\tilde{\bt},\ba^{(m+1)})}{\Pr(\bt^{(m)},\ba^{(m+1)})}
\Bigg\}.
$$

\item[$\circ$] Set \ \ 
$
\bt^{(m+1)} = \left\{ \begin{array}{l l}
\tilde{\bt} & \qquad \text{if} \ \  \gamma_{\bt} \ge u \sim \text{Uniform}(0,1), \\
\bt^{(m)} &  \qquad \text{otherwise}.
\end{array} \right.
$
\end{itemize}

 \end{itemize}

\medskip
\ \ \ \ {\bf end for}

\medskip
\STATE {\bf 4.} {\it Output}: return $\{ \ba^{(m)} \}_{m=1}^M$ and $\{ \bt^{(m)} \}_{m=1}^M$.

\end{algorithmic}
\end{algorithm}

A few remarks about the choice of prior and proposal distributions, as well as convergence monitoring in Algorithm \ref{ALG_MG}, follow. 

\paragraph{I. Prior distributions.} 
The choice of prior distributions for the parameters $\bt$ and $\ba$ should be based on any existing knowledge or assumptions about the system. Priors can be non-informative (e.g., uniform distributions) to reflect a lack of strong prior knowledge. They can also be informative (e.g., Gaussian distributions) to incorporate specific prior beliefs. For instance, we know that the hyper-parameters $\ba$ are all positive, or we may have information about the range of model parameters $\bt$; see Section \ref{sec:quantum_char} for examples. Additionally, by assuming that the model parameters and hyper-parameters are independent, we can express the joint prior as $Pr(\bt,\ba)=Pr(\bt)Pr(\ba)$. Selecting appropriate priors is essential to ensure that the posterior distribution accurately reflects both the data and any prior knowledge.

\paragraph{II. Proposal distributions.} The efficiency of the Metropolis sampling steps heavily depends on the choice of proposal distributions. These distributions determine the candidate values for the parameters in each iteration. A well-chosen proposal distribution should allow the algorithm to explore the parameter space efficiently while maintaining an adequate acceptance rate. Common choices include Gaussian proposals centered at the current parameter values, with the variance controlling the step size. Adaptive proposal schemes can also be used to dynamically adjust the step size during the sampling process; see, e.g., \cite{haario2001adaptive}. 
In Section \ref{sec:quantum_char}, we use uniform random walk proposals to generate new samples $\tilde{\ba}$ and $\tilde{\bt}$ from current samples $\ba^{(m)}$ and $\bt^{(m)}$:
\begin{equation}\label{proposal_random_walk}
\tilde{\ba} \sim \text{Uniform} (\ba^{(m)} - {\bf r}_{\ba}/2, \,
\ba^{(m)} + {\bf r}_{\ba}/2), \qquad 
\tilde{\bt} \sim \text{Uniform} (\bt^{(m)} - {\bf r}_{\bt}/2, \,
\bt^{(m)} + {\bf r}_{\bt}/2),
\end{equation}
where ${\bf r}_{\ba}$ and ${\bf r}_{\bt}$ are two support vectors, to be selected so that the proposal distributions are neither too wide, nor too narrow. 
If a proposal gives a new sample outside the support of the prior, we may simply discard the sample and draw another sample from the proposal.  
It is to be noted that the choice \eqref{proposal_random_walk} induces symmetric proposals, i.e., 
$q_{\ba}(\ba^{(m)},\tilde{\ba}) = q_{\ba}(\tilde{\ba},\ba^{(m)})$ 
and 
$q_{\bt}(\bt^{(m)},\tilde{\bt}) = q_{\bt}(\tilde{\bt},\bt^{(m)})$. 
If, instead of symmetric proposals, we opted for non-symmetric proposals, then we would need to multiply the acceptance ratios in $\gamma_{\ba}$ and $\gamma_{\bt}$ by $q_{\ba}(\ba^{(m)},\tilde{\ba})/ q_{\ba}(\tilde{\ba}, \ba^{(m)})$ and $q_{\bt}(\bt^{(m)},\tilde{\bt})/ q_{\bt}(\tilde{\bt}, \bt^{(m)})$, respectively. 
This would in turn require extra evaluations of the proposals.

\paragraph{III. Convergence Monitoring.} 
Monitoring the convergence of the Markov chain is crucial to ensure that the samples are drawn from the target posterior distribution. Various diagnostic tools, such as trace plots, effective sample size estimates, and convergence statistics like the Gelman-Rubin diagnostic, can be used to assess whether the chain has converged. It is important to run the chain long enough and potentially use multiple chains to verify convergence. 

\subsection{Gaussian process predictions}
\label{sec:prediction}
Once the noisy measurements ${\bf y}_n = (y_1, \dotsc, y_n)$ at a set of {\it training} points ${\bf t} = (t_1, \dotsc, t_n)$ are collected, and independent samples $\{ \bt^{(m)} \}_{m=1}^M$ and $\{ \ba^{(m)} \}_{m=1}^M$ from the posterior distribution $\Pr(\bt, \ba | {\bf y}_n)$ are obtained, predictions can be made at a new set of {\it test} points ${\bf t}^* = (t_1^*, \dotsc, t_{n^*}^*)$. Note that the number $n^*$ of test points may differ from the number $n$ of training points. 
To make predictions, we define the vector ${\bf g}^*$ as:
$$
{\bf g}^* = (g(t_1^*), \dotsc, g(t_{n^*}^*) ), \qquad g(t_i^*) := f(t_i^*; \bt) + \delta(t_i^*), \qquad i=1, \dotsc, n^*.
$$
Here, ${\bf g}^* \in {\mathbb R}^{n^*}$ is a random vector dependent on $\bt$. Samples of ${\bf g}^*$ can be generated by computing $f$ using parameter samples from the posterior and sampling the zero-mean Gaussian process $\delta$ with a deterministic covariance kernel $\hat{\kappa}$, as defined by \eqref{specific_covariance}. The kernel $\hat{\kappa}$ is characterized by a deterministic variance $\hat{\sigma}_{\delta}$ and a deterministic time scale $\hat{\tau}$, which are obtained from the posterior sample mean or mode of the hyper-parameters. 
Specifically, for a given posterior parameter sample $\bt^{(m)}$, we compute ${\bf f}^*(\bt^{(m)}) = (f(t_1^*; \bt^{(m)}), \dotsc, f(t_{n^*}^*; \bt^{(m)}))$ and add a sample ${\bf d}^{(m)}$ drawn from a multivariate normal distribution with zero mean and covariance matrix $\hat{K}^{*} = [\hat{\kappa}(t_i^*,t_j^*)] \in {\mathbb R}^{n^* \times n^*}$:
$$
{\bf g}^{*(m)} = {\bf f}^*(\bt^{(m)}) + {\bf d}^{(m)}, \qquad {\bf d}^{(m)} \sim {\mathcal N}({\bf 0}, \hat{K}^*).
$$

\subsection{Related works and novelty of the marginal-likelihood approach}

Gaussian processes provide a flexible probabilistic modeling framework but face computational challenges due to the \( O(n^3) \) complexity of inverting the \( n \times n \) covariance matrix. Numerous methods mitigate this burden, often via approximations that trade accuracy for efficiency. 
Low-rank approximations, such as the Nystr{\"o}m method~\cite{williams2001using} and sparse variational GPs~\cite{titsias2009variational}, approximate the covariance matrix \( K \) by selecting a subset of inducing points or optimizing a variational representation. Similarly, random feature expansions~\cite{rahimi2007random} approximate the kernel function using a Fourier basis, transforming GP inference into parametric regression. Localized strategies, including mixtures of experts~\cite{rasmussen2002infinite} and pseudo-point approximations~\cite{snelson2007local}, partition the input space into independent GP sub-models, sacrificing global coherence and introducing boundary artifacts. 
Structured kernel methods, such as Kronecker decompositions for grid-structured data~\cite{saatci2012scalable} and Toeplitz approximations for stationary kernels~\cite{wilson2014fast}, exploit algebraic structure but impose restrictions on the kernel and input geometry. Active subspace techniques~\cite{constantine2015active, tripathy2016gaussian, zahm2020gradient} reduce dimensionality by identifying dominant directions via gradients, but their reliance on derivative calculations makes them costly and less robust in high-dimensional, noisy settings. Even foundational stabilization techniques, such as Cholesky factorization with jitter~\cite{Rasmussen_Williams:06}, ensure numerical invertibility by perturbing the covariance matrix but introduce distortions to the prior.

In contrast, our marginalized likelihood strategy addresses covariance degeneracy directly, without resorting to approximations or artificial perturbations. By analytically integrating out the degenerate subspace, we preserve the original kernel structure, ensuring both numerical stability and computational efficiency. Unlike low-rank methods, our approach does not require explicit kernel approximations via inducing points or variational optimization. Unlike local approximation techniques, we retain a unified global model. Unlike structured kernel methods, our framework remains kernel-agnostic. And unlike active subspaces, we avoid expensive gradient computations, instead using eigendecomposition to identify dominant directions—an approach conceptually aligned with Principal Component Analysis (PCA) \cite{hotelling1933}. However, we go beyond PCA by marginalizing the likelihood over the degenerate subspace, effectively integrating out directions that contribute negligible information. 
This method bridges the gap between exact inference and scalability, preserving the interpretability of classical GPs while overcoming the limitations of existing approximation-based techniques. 
As demonstrated in Section \ref{sec:quantum_char}, our strategy provides a stable and important addition to the GP computational toolkit, delivering significant computational gains without compromising model fidelity.

\section{Application to quantum characterization}
\label{sec:quantum_char}

In this section, we discuss the deployment of the proposed Bayesian approach to the characterization of a superconducting quantum device using population measurements.

\subsection{Quantum characterization: background and literature review}

Quantum characterization is the process of estimating quantum device parameters, such as transition frequencies and decoherence times. 
This process can be formulated as an inference problem: given a parametric quantum dynamical model and a set of quantum measurements, the goal is to infer the quantum model parameters. Quantum characterization is a crucial step for quantum control, which is essential for implementing robust quantum operations; see e.g., \cite{Glaser_etal:2015,Koch_etal:2022}. For a recent and comprehensive overview of quantum characterization, we refer the reader to \cite{Review_quantum_chhar:2024}.

Despite recent advancements, current Noisy Intermediate-Scale Quantum (NISQ) systems face significant challenges with limited coherence, primarily due to environmental noise and gate infidelities caused by drift in device parameters \cite{Bharti_etal:2022}. Additional sources of uncertainty in quantum dynamical modeling include state preparation and measurement (SPAM) errors, variability in amplifier gain, and model-form uncertainty stemming from the limitations of quantum dynamical models like the Schr{\"o}dinger and Lindblad equations. Factors such as unknown Hamiltonian or Lindbladian operators, non-Markovian decoherence mechanisms, and violations of local conservation laws and thermalization further contribute to model-form errors \cite{Tupkary_etal:2022}. These complexities make the task of quantum characterization particularly challenging.

Conventional methods for determining quantum device parameters typically rely on deterministic curve fitting techniques and frequency domain analysis applied to a parametric Hamiltonian model that describes the device's dynamics; see e.g., \cite{Peterer_etal:2015,Krantz_etal:2019,Wittler_etal:2021}. However, these methods do not consider the inherent uncertainty in current NISQ systems. 
To address this limitation, various Bayesian quantum characterization and Hamiltonian learning approaches have been developed; see e.g., \cite{Wiebe_etal:2014a,Wiebe_etal:2014b,Wiebe_etal:2015,Wang_etal:2017,Ralph_etal:2017,Evans_etal:2019,Bennink_etal:2019,Gentile_etal:2021,Peng_etal:2023}. Bayesian methods account for uncertainty in quantum device parameters by providing posterior probability distributions rather than deterministic values. These posteriors can guide the design of quantum devices and support the development of more robust controls through risk-neutral optimization \cite{Shapiro_etal:2021}. 
However, existing Bayesian techniques primarily address experimental uncertainty and overlook model-form uncertainty by assuming the quantum dynamical model to be accurate.

In this section, we present a quantum characterization technique that leverages the proposed marginal likelihood method to account for both experimental and model-form uncertainty, thereby enhancing the predictive capability of existing quantum models.

\subsection{Quantum measurements }
\label{sec:Ramsey_exp}

The experiments in this study are performed on a tantalum-based superconducting transmon device \cite{Place_etal:2021} at LLNL. In principle, a transmon can support many energy levels. However, on this system only the three lowest levels, corresponding to the states $|0\rgl$, $|1\rgl$ and $|2\rgl$, can be reliably measured. In the following, the transmon device will be referred to as the ``qudit". For simplicity, we limit our discussion to the Ramsey experimental protocol, which is designed to determine the transition frequencies $\omega_{k,k+1}$ and the combined decoherence time scales $T^{*}_{2,k}$~\cite{Tempel_AspuruGuzik:2011}, with $k=0,1, \ldots$. Here, the frequency $\omega_{k,k+1}$ corresponds to transition between the states $|k\rgl$ and $|k+1\rangle$. Because only the first three states ($|0\rgl$, $|1\rgl$ and $|2\rgl$) can be accurately measured on the qudit, we limit the experiments to determine the $0 \leftrightarrow 1$ and $1 \leftrightarrow 2$ transition frequencies as well as the corresponding dephasing times.

During each application of the Ramsey $k \leftrightarrow k+1$ protocol, we first apply a series of $\pi$-pulses to prepare the qudit in state $|k\rgl$, followed by a detuned $\frac{\pi}{2}$-pulse to bring the device into a superposition of states $|k\rgl$ and $|k+1\rgl$. The system is then evolved freely (without applying any control pulses) during the dark time $t_{\textrm{dark}}$, after which a second detuned $\frac{\pi}{2}$-pulse is applied. Finally, the resulting state of the system is measured. The measurement process causes the quantum state to collapse to a classical state, in accordance with the principles of quantum mechanics. The experiment is repeated 1,000 times (i.e., 1,000 shots) for each dark time and the frequency of each state is recorded to estimate the probability (or population) of that state. See the appendix of~\cite{Peng_etal:2023} for further details. 

\begin{figure}[thb]
  \begin{center} 
    \begin{tikzpicture}
    \node (ImgRamsey){\includegraphics[width=0.47\textwidth]{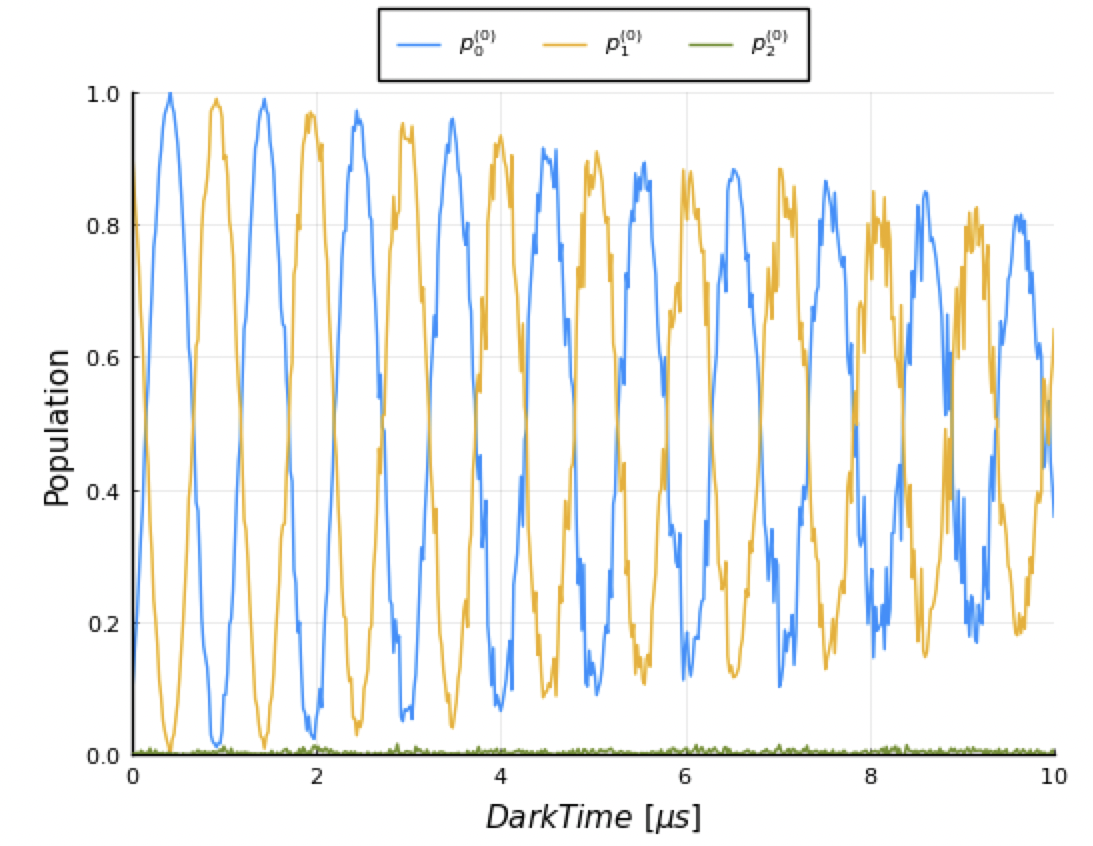}};
    \node(Ramsey)[below=of ImgRamsey,yshift=+1.2cm]{\scriptsize$\frac{\pi_{0,1}}{2}\xrightarrow[\textrm{free evolution}]{t_{\textrm{dark}}}\frac{\pi_{0,1}}{2}$};
    \node(Ramsey)[left=of Ramsey,yshift=-0.05cm,xshift=1.0cm]{\scriptsize Protocol:};
  \end{tikzpicture}
   \begin{tikzpicture}
    \node (ImgRamsey12){\includegraphics[width=0.47\textwidth]{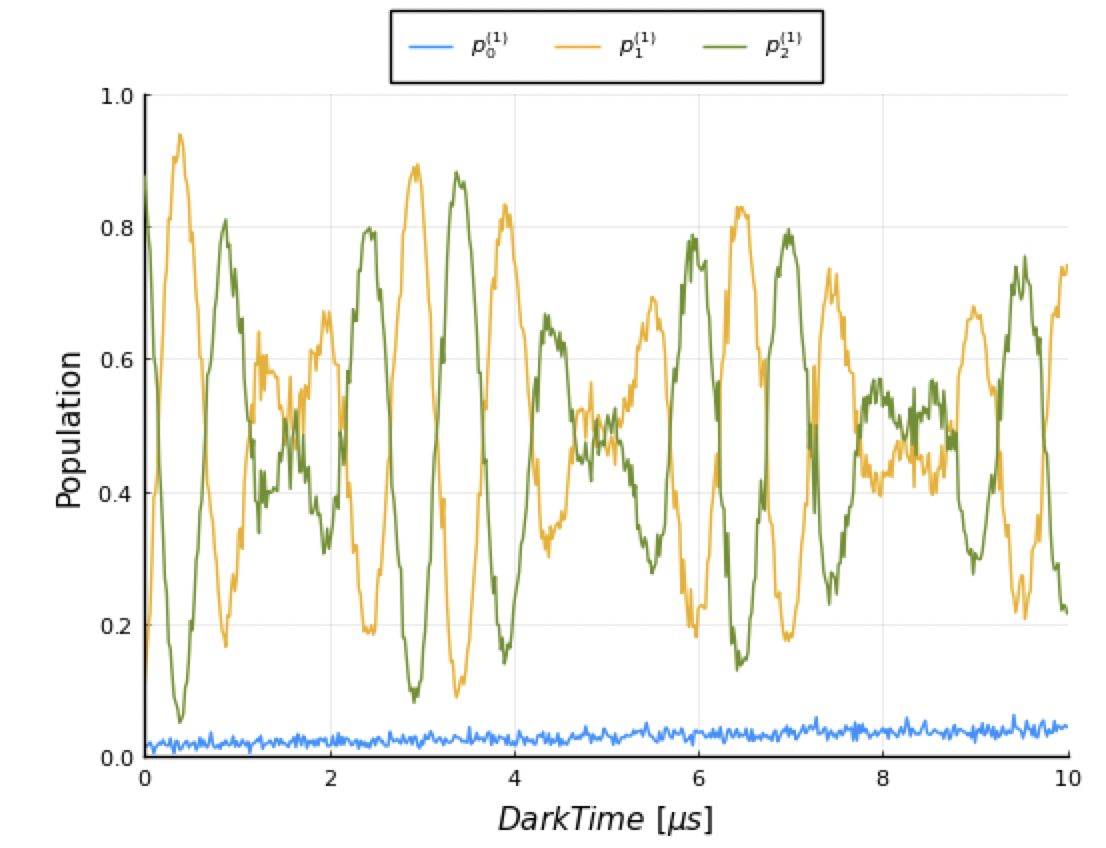}};
    \node(Ramsey12)[below=of ImgRamsey12,yshift=+1.2cm]{\scriptsize$\pi_{0,1}\rightarrow\frac{\pi_{1,2}}{2}\xrightarrow[\textrm{free evolution}]{t_{\textrm{dark}}}\frac{\pi_{1,2}}{2}$};
  \end{tikzpicture}
    \caption{Populations as functions of dark time for Ramsey $0 \leftrightarrow 1$ (left) and $1 \leftrightarrow 2$ (right) experiments. In the protocol, $\pi_{k,k+1}$ and $\frac{\pi_{k,k+1}}{2}$ denote $\pi$ and $\frac{\pi}{2}$ pulses for the $k\leftrightarrow k+1$ transition.}
    \label{fig:exps}
 \end{center}
\end{figure}

Each Ramsey experiment is performed for $n=500$ dark times, discretized on a uniform grid with a step size of $\Delta t =$ \SI{20}{\nano\second}. To setup the Ramsey experiment, the drive frequency (which determines the amount of detuning) is chosen as the estimated transition frequency from a standard calibration procedure, reduced by a \SI{1}{\mega\hertz} nominal detuning. This procedure resulted in the drive frequencies $\omega_d/2\pi = 3.4476698$ \SI{}{\giga\hertz} and $\omega_d/2\pi = 3.2392576$ \SI{}{\giga\hertz}, for the $0 \leftrightarrow 1$ and $1 \leftrightarrow 2$ Ramsey experiments, respectively; see, e.g.,~\cite{qiskit_url} for details. 
Results from the Ramsey experiments are presented in Figure \ref{fig:exps}, showing the measured populations as functions of dark time. 

Here, the Ramsey measurements are distinctly corrupted by noise, and the noise has a complex structure. In the $0\leftrightarrow 1$ case, the noise increases as the dark time increases, and in the $1\leftrightarrow 2$ case, the noise increases largely in the phase-flip regions around dark times 1.5 \SI{}{\micro\second}, 5.0 \SI{}{\micro\second}, and 8.5 \SI{}{\micro\second}. 
We attribute these beatings and phase flips to random parity events \cite{Riste_etal:2013}. 
These random events occur on a time scale of milliseconds and alternate the transition frequencies of the device through sudden changes in the charge parity. 
Because the population data oscillates with detuning frequency $\Delta_k=\omega_{k,k+1}-\omega_d$, it is expected that the Fourier spectrum of the population data will exhibit a maximum at the detuning frequency. As reported by Peng et al.~\cite{Peng_etal:2023} (also see~\cite{Tennant_etal:2022}), a single peak is observed in the $0\leftrightarrow 1$ spectrum, but two distinctive peaks are present in the $1\leftrightarrow 2$ case. Given these results, we conclude that changes in charge parity (charge dispersion) is significant only in the latter case. 

\subsection{Simulation models }

To take Markovian interactions with the environment into account we model the oscillatory dynamics of the qudit state using Lindblad's master equation~\cite{Lindblad:1976,Nielsen_Chuang:2000}, which is a system of ordinary differential equations posed as an initial-value problem:
\begin{align}
    \dot{\rho}(t) &= -i\left(H(t)\rho(t) - \rho(t) H(t)\right)
    +\sum_{j=1}^{N^2 - 1} \left( {\cal L}_{j} \rho(t) {\cal L}_{j}^\dagger -
\frac{1}{2}\left( {\cal L}_{j}^\dagger{\cal L}_{j}\rho(t) + \rho(t) {\cal L}_{j}^\dagger{\cal L}_{j} \right) \right),\ t>0,
\label{eq:lindblad}
\\ 
\rho(0)&=\rho_0.
\label{eq:lindblad_0}
\end{align}
Here, $\rho=\rho^\dagger$ is the density matrix, $H=H^\dagger$ is the Hamiltonian and ${\cal L}_j$ denotes a decoherence operator. These operators are all in $\mathbb{C}^{N\times N}$, with $N$ being the size of the Hilbert space. The diagonal elements of the density matrix represent the population of the states, while the off-diagonal elements capture the coherence between states in the system. Only the three lowest energy levels can currently be reliably measured on the qudit, but to reduce artificial effects from truncation of the Hilbert space, we add a guard level and include the four lowest energy levels of the system in our modeling. Hence, we set $N=4$. 
The initial condition \eqref{eq:lindblad_0} is determined by the qudit's initialization. For example, in a Ramsey experiment, where the qudit is initially prepared in the ground state $|0\rgl$, the initial density matrix, $\rho_0$, is defined as a diagonal matrix with a 1 in its first diagonal element and 0 elsewhere.

The Hamiltonian operator in Lindblad's equation \eqref{eq:lindblad} is of the form $H(t)=H_s+H_c(t)$, 
with the system Hamiltonian $H_s$ and the control Hamiltonian $H_c(t)$ given by
\begin{equation*}
H_{s} =\left(
 \begin{array}{cccc}
0 & 0 & 0 & 0 \\
0 & \omega_{0,1} & 0 & 0 \\
0 & 0 & \omega_{0,1} + \omega_{1,2} & 0\\
0 & 0 &  0 & \omega_{0,1} + \omega_{1,2}+\omega_{2,3} \\
\end{array}
\right),
\end{equation*}
and 
\begin{equation*}
    H_c(t) = h(t)(a+a^\dagger), 
    \qquad 
    h(t) =  2\, I(t) \, \cos(\omega_d t) + 2 \, Q(t) \, \sin(\omega_dt).
\end{equation*}
Here, $a$ and $a^\dagger$ are the lowering and raising operators, $\omega_d$ is the angular drive frequency, and $h(t)$ is the control function given in terms of two slowly varying envelope functions $I$ and $Q$, referred to as the in-phase and the quadrature components, respectively. 
These components are typically used as input signals to an IQ-mixer that, in turn, generates the control signal $h(t)$ that is sent to the qudit.

We consider two decoherence operators\footnote{Although with $N=4$, there could theoretically be up to $N^2-1 = 15$ decoherence operators; the two operators considered here are the most important ones.} in Lindblad's equation \eqref{eq:lindblad}: the decay operator $\mathcal{L}_1$ and the dephasing operator $\mathcal{L}_2$, defined as
\begin{align*}
{\cal L}_{1} &= \begin{pmatrix}
0 & \sqrt{\gamma_{1,1}} & 0 & 0 \\
0 & 0 & \sqrt{\gamma_{1,2}} & 0  \\
0 & 0 & 0 & \sqrt{\gamma_{1,3}} \\
0 & 0 & 0 & 0
\end{pmatrix}, 
\qquad
{\cal L}_{2} = \begin{pmatrix}
0 & 0 & 0 & 0\\
0 &\sqrt{\gamma_{2,1}}& 0 & 0\\
0 & 0 & \sqrt{\gamma_{2,2}} & 0 \\
0 & 0 & 0 & \sqrt{\gamma_{2,3}}
\end{pmatrix}.
\end{align*}
The parameter $\gamma_{1,k}$ is the decay rate for state $|k\rgl$, with corresponding decay time $T_{1,k}=1/\gamma_{1,k}$. The dephasing rates $\gamma_{2,k}$ are related to the pure dephasing times $T_{2,k}$ through the relations, $\gamma_{2,0}=0$, $\sqrt{\gamma_{2,k}}=\sqrt{\gamma_{2,k-1}}+\sqrt{2/T_{2,k}}, \, k=1,2,3$; see the appendix of~\cite{Peng_etal:2023_mathematical} for details. The decay and pure dephasing time scales are related to the combined decoherence time scales through $1/T^{*}_{2,k} = 1/(2\,T_{1,k}) + 1/T_{2,k}$; see~\cite{Tempel_AspuruGuzik:2011}.

The equations above are stated in the laboratory frame of reference. However, as the transition and drive frequencies are high, typically in the \SI{}{\giga\hertz} range, numerical simulations can be computationally expensive. 
We slow down the time scales in the density matrix by applying the rotating wave approximation (RWA) and transforming \eqref{eq:lindblad} into a frame rotating with the angular frequency $\omega_d$; see for example~\cite{Petersson_etal:2020}. 
In the present work, we only use control functions that are piece-wise constant in the rotating frame of reference. Because the Hilbert space is low-dimensional, a highly efficient approach for integrating Lindblad's master equation is through matrix exponentiation.

The random parity event in the $1\leftrightarrow 2$ transition frequency is modeled by
$$
    \omega_{1,2} = \bar{\omega}_{1,2} + p\, \epsilon_{1,2}, \quad p\in\{-1,1\}, 
$$
where $\bar{\omega}_{1,2}$ is the average $1\leftrightarrow 2$ transition frequency, $\epsilon_{1,2}$ is the charge dispersion and $p\in\{-1,1\}$ is a discrete random variable, called the parity, taking values $\pm 1$ with equal probability \cite{Riste_etal:2013}. We define the frequencies corresponding to the positive and negative parities as
\[
\omega^\pm_{1,2}=\bar{\omega}_{1,2}\pm\epsilon_{1,2}.
\]
Since $p$ is assumed to be a discrete random variable with zero mean, the average $1\leftrightarrow 2$ transition frequency and the charge dispersion are \mbox{$\bar{\omega}_{1,2} = \frac{1}{2}(\omega_{1,2}^++\omega_{1,2}^-)$} and $\epsilon_{1,2}=\frac{1}{2}(\omega_{1,2}^+-\omega_{1,2}^-)$, respectively.

Parity events have been reported to occur on a time scale of milliseconds~\cite{Riste_etal:2013}. This time scale is much longer than the duration of a single shot of the experiments, which typically only requires a few microseconds. Although a parity event could occur during a single shot, the disparate time scales indicate that this would be unlikely. On the other hand, the state population is often measured by averaging over $1,000$ (or more) shots. In our experiments the wait time between successive shots for the same dark time varies between \SI{0.05}{\milli\second} and \SI{0.1}{\milli\second}. As a result it is likely that about half of the shots are performed for each parity. To account for both parities we solve Lindblad's master equation twice, once with $\omega_{1,2}^+$ and once with $\omega_{1,2}^-$, resulting in the density matrices $\rho^+$ and $\rho^-$, respectively. The average of these density matrices is then used in the characterization, described below.

\subsection{Marginal-Likelihood quantum characterization}
\label{sec:GP_char}

Let the unknown parameters of Linblad's master equation \eqref{eq:lindblad} consist of the transition frequencies and the pure dephasing time scales,
$$
{\bt} = (\omega_{0,1},\omega_{1,2}^-,\omega_{1,2}^+, T_{2,1}, T_{2,2}). 
$$
Since the $T_1$-decay times cannot be characterized by Ramsey experiments, we assume they are already determined by some other protocol, e.g., energy decay experiments. Here, we use the fixed values $T_{1,1}=258.39$ $\mu$s and $T_{1,2} = 100.79$ $\mu$s from \cite{Peng_etal:2023}. 

In the following, we let $p^{(k)}_s(t_i)$ be the measured population of state $s \in \{0,1,2\}$, recorded at dark time $t_i$ in the Ramsey $k \leftrightarrow k+1$ experiment, with $k=0,1$. We collect measurements corresponding to $n=500$ dark times $t_i=i \, \Delta t$, with $\Delta t =$ \SI{20}{\nano\second} and $i=1,2,\ldots,n$. 
Also, we let $\hat{p}^{(k)}_s(t_i;\bt)$ denote the simulated population of state $s \in \{0,1,2\}$, 
obtained by solving the Lindblad initial-value problem \eqref{eq:lindblad}-\eqref{eq:lindblad_0} for the Ramsey $k \leftrightarrow k+1$ experiment with parameters $\bt$. Specifically, $\hat{p}^{(k)}_s(t_i;\bt)$ corresponds to the $s$-th diagonal element of the density matrix at the completion of the simulated Ramsey protocol, with dark time $t_i$.

Following the Bayesian approach in Section \ref{sec:general} and analogous to the noise-discrepancy model \eqref{data_model}, we consider the following relationships between the experimental and simulated measurements for the two Ramsey experiments:
\begin{align}
p^{(0)}_s (t_i) &= \hat{p}^{(0)}_s(t_i;\bt) + \delta^{(0)}(t_i) + \varepsilon^{(0)}, 
\qquad s=0,1,2, \qquad i=1,2,\ldots,n, \label{model_R0}\\
p^{(1)}_s (t_i) &= \hat{p}^{(1)}_s(t_i;\bt) + \delta^{(1)}(t_i) + \varepsilon^{(1)}, 
\qquad s=0,1,2, \qquad i=1,2,\ldots,n, \label{model_R1}
\end{align}
where
\begin{equation}\label{cov_quantum}
\varepsilon^{(k)} \sim \mathcal{N}(0,\sigma_{\varepsilon,k}^2), 
\quad 
\delta^{(k)}(t) \sim {\mathcal {GP}}(0, \kappa^{(k)}(t,t')), 
\quad 
\kappa^{(k)}(t,t') = \sigma_{\delta,k}^2 \, \exp \left( -\frac{ |t - t'|^{\gamma}}{2 \, \tau_k^\gamma}\right),
\quad k=0,1.
\end{equation}
We note that since the Ramsey $k \leftrightarrow k+1$ measurements have different noise-discrepancy structures for different $k \in \{0,1\}$, as can be seen in Figure \ref{fig:exps}, we consider different noise-discrepancy parameters $\ba^{(0)} = (1/\sigma_{\varepsilon,0}^2, 1/\sigma_{\delta,0}^2, \tau_0)$ and $\ba^{(1)} = (1/\sigma_{\varepsilon,1}^2, 1/\sigma_{\delta,1}^2, \tau_1)$ across the two Ramsey experiments. 
Within each Ramsey experiment, however, we assume that the noise and model discrepancy have the same covariance structures. The power in the exponent, $\gamma >0$, determines the smoothness of the Gaussian process and may also be considered as a hyper-parameter. In the following, we will however only perform numerical experiments with the fixed value $\gamma = 1$; see Section \ref{sec:ex2}.

We further note that the three components $p_0^{(k)}(t), p_1^{(k)}(t), p_2^{(k)}(t)$ represent the probabilities that a measurement taken at dark time $t$ results in one of the three outcomes $\{0,1,2\}$. As a result, they (approximately) sum up to one at every dark time. This implies that in each Ramsey $k \leftrightarrow k+1$ experiment, one of the three components may not be informative and hence may be excluded from the data set. 
For this reason, we exclude $p_2^{(0)}$ in the Ramsey $0 \leftrightarrow 1$ model \eqref{model_R0} and $p_0^{(1)}$ in the Ramsey $1 \leftrightarrow 2$ model \eqref{model_R1}, noting that they are much smaller than the other probability components; see Figure \ref{fig:exps}. 

Now, setting
\begin{align*}
{\bf p}_s^{(0)} &= \{ p_s^{(0)}(t_i) \}_{i=1}^n, \qquad
\hat{\bf p}_s^{(0)}(\bt) = \{ \hat{p}_s^{(0)}(t_i; \bt) \}_{i=1}^n, \qquad
\Sigma(\ba^{(0)}) = K(\sigma_{\delta,0},\tau_0) + \sigma_{\varepsilon,0}^2 \, I_n, \quad s=0,1, \\    
{\bf p}_s^{(1)} & = \{ p_s^{(1)}(t_i) \}_{i=1}^n, \qquad
\hat{\bf p}_s^{(1)}(\bt) = \{ \hat{p}_s^{(1)}(t_i; \bt) \}_{i=1}^n, \qquad
\Sigma(\ba^{(1)}) = K(\sigma_{\delta,1},\tau_1) + \sigma_{\varepsilon,1}^2 \, I_n, \quad s=1,2,
\end{align*}
and assuming that our data set, say $D$, consists of four independent discrete time series, 
$$
D = \{ D^{(0)}, \, D^{(1)} \} \in {\mathbb R^{4 n}}, \qquad 
D^{(0)} = \{ {\bf p}_0^{(0)}, \, {\bf p}_1^{(0)} \} \in {\mathbb R^{2 n}}, \qquad  
D^{(1)} = \{ {\bf p}_1^{(1)}, \, {\bf p}_2^{(1)} \} \in {\mathbb R^{2 n}},
$$
we arrive at the likelihood, 
\begin{equation}\label{likelihood_quantum}
\Pr (D | \bt, \ba) =  \Pr (D^{(0)} | \bt, \ba^{(0)}) \, \Pr (D^{(1)} | \bt, \ba^{(1)}),
\end{equation}
where
\begin{align*}
\Pr (D^{(0)} | \bt, \ba^{(0)}) &= (2 \pi)^{-n} \, |\Sigma(\ba^{(0)})|^{-1} \, 
\prod_{s=0}^1 
 \exp \left( - \frac{1}{2} ({\bf p}_s^{(0)} - {\bf p}_s^{(0)}(\bt))^{\top} \Sigma^{-1}(\ba^{(0)}) \, ({\bf p}_s^{(0)} - {\bf p}_s^{(0)}(\bt)) \right), \\
\Pr (D^{(1)} | \bt, \ba^{(1)}) &= (2 \pi)^{-n} \, |\Sigma(\ba^{(1)})|^{-1} \, 
\prod_{s=1}^2 
 \exp \left( - \frac{1}{2} ({\bf p}_s^{(1)} - {\bf p}_s^{(1)}(\bt))^{\top} \Sigma^{-1}(\ba^{(1)}) \, ({\bf p}_s^{(1)} - {\bf p}_s^{(1)}(\bt)) \right).
\end{align*}

Assuming that $\bt$ and $\ba$ are independent, we express the joint prior as $Pr(\bt,\ba)=Pr(\bt)Pr(\ba)$. We use a multivariate uniform prior for $\bt$ based on the deterministic characterization performed in \cite{Peng_etal:2023}. Specifically, we set 
\begin{equation}\label{uniform_prior}
\Pr(\bt) = \prod_{i=1}^5 \Pr(\theta_i), \qquad \theta_i \sim \text{Uniform}(l_i,u_i),
\end{equation}
with the lower and upper bounds
$$
\arraycolsep=10pt
\begin{array}{lll}
l_1 / 2\pi = \bar{\omega}_{0,1}/2\pi - 10^{-3} \, \text{GHz}, 
\qquad 
& u_1 / 2\pi = \bar{\omega}_{0,1}/ 2\pi + 10^{-3} \, \text{GHz}, 
\qquad 
& \bar{\omega}_{0,1}/2\pi = 3.448646 \, \text{GHz}, \\
l_2 / 2\pi = \bar{\omega}^-_{1,2}/ 2\pi - 10^{-3} \, \text{GHz}, \qquad  
& u_2 / 2\pi = \bar{\omega}^-_{1,2}/ 2\pi + 10^{-3} \, \text{GHz}, \qquad  
& \bar{\omega}^-_{1,2}/2\pi = 3.240105 \, \text{GHz}, \\ 
l_3 / 2\pi = \bar{\omega}^+_{1,2}/ 2\pi - 10^{-3} \, \text{GHz}, \qquad 
& u_3 / 2\pi = \bar{\omega}^+_{1,2} / 2\pi + 10^{-3} \, \text{GHz}, \qquad 
& \bar{\omega}^-_{1,2}/2\pi = 3.240403 \, \text{GHz}, 
\end{array}
$$
$$
\arraycolsep=20pt
\begin{array}{lll}
l_4 = \bar{T}_{2,1} - 5 \, \mu\text{s}, 
\hskip 1cm
& u_4 = \bar{T}_{2,1} + 5 \, \mu\text{s}, 
\hskip 1cm
& \bar{T}_{2,1} = 13.07 \, \mu\text{s}, \\
l_5 = \bar{T}_{2,2} - 1.5 \, \mu\text{s},  
\hskip 1cm
& u_5 = \bar{T}_{2,2} + 1.5 \, \mu\text{s}, 
\hskip 1cm
& \bar{T}_{2,2} = 2.73 \, \mu\text{s}.
\end{array}
$$
We also use uniform priors for the hyper-parameters,  
\begin{equation}\label{uniform_prior_hyper}
\Pr(\ba^{(0)}) = \Pr(\ba^{(1)}) = \prod_{i=1}^3 \Pr(\alpha_i), \qquad \alpha_1, \alpha_2 \sim \text{Uniform}(1,10^4), \qquad \alpha_3 \sim \text{Uniform}(0.1,10),
\end{equation}
ensuring that they will remain positive while allowing them to take on large values.

We note that the assumption of independence between $\bt$ and $\ba$ arises from their fundamentally distinct roles: $\bt$ governs intrinsic dynamics, while $\ba$ pertains to noise and error characteristics, with no prior evidence suggesting direct dependency. In the large-data regime considered here, the likelihood dominates the posterior, minimizing the impact of any inaccuracies in this assumption.

With the full likelihood \eqref{likelihood_quantum} and the priors \eqref{uniform_prior}-\eqref{uniform_prior_hyper} in hand, we can apply the proposed sampling method of Sections \ref{sec:marginal_likelihood} and Algorithm \ref{ALG_MG} to perform inference and prediction.

\subsection{Predictive capability of the proposed approach}
\label{sec:results}

We now demonstrate the predictive capability of the proposed Bayesian quantum characterization approach in comparison with a classical Bayesian approach. Specifically, we consider two cases: 
\begin{itemize}
\item[{\bf 1}.] Bayesian characterization without model discrepancy;

\item[{\bf 2}.] Bayesian Gaussian process characterization including model discrepancy. 
\end{itemize}

We will show that the second approach outperforms the first, not only in computing the quantum system's response at the recorded dark time points, but also in predicting the system's response at interpolatory time points.

In both cases, we run Algorithm \ref{ALG_MG} for $M=2 \times 10^4$ iterations and remove the first half of the samples, known as the burn-in period. We also use a thinning period of 2, that is, we discard every other sample in the chain to reduce the correlation between consecutive samples. As a result, we get a total of $5000$ Markov chain samples from parameter posteriors. 
We use the symmetric proposals in \eqref{proposal_random_walk} with the supports ${\bf r}_{\bt} = (2 \pi \times 10^{-6}, 2 \pi \times 10^{-6}, 2 \pi \times 10^{-6},0.2,0.1)$ and ${\bf r}_{\ba^{(0)}} = {\bf r}_{\ba^{(1)}} = (8,8,0.05)$. 
As a standard test, for each approach, we run multiple Markov chains with different initial values of parameters and monitor the trace plots to ensure a good mixing and convergence of the chains and acceptance rates of $20-25 \%$. 
The computational details specific to each case are given below.

\subsubsection{Approach 1: Bayesian characterization without model discrepancy}
\label{sec:ex1}

We use Algorithm \ref{ALG_MG} with the likelihood function \eqref{likelihood_quantum}, but exclude the Gaussian process covariance matrix $K$ from the likelihood's covariance matrix $\Sigma$. That is, we consider the covariance only due to experimental noise: $\Sigma(\ba^{(k)}) = \sigma_{\varepsilon,k}^2 I_n$, with $k=0,1$. 
Note that in this case, there is no need to compute the dominant eigenvalues and corresponding eigenvectors.

The posterior densities and the mean and standard deviation of the five main parameters $\bt$ can be found in Figure \ref{fig:posteriors} and Table \ref{tab:stat}, respectively. The posterior densities are obtained by normalizing their corresponding histograms generated by the Markov chain samples. 
The mean of hyper-parameters, i.e., the standard deviations (${\sigma}_{\varepsilon,0}$,${\sigma}_{\varepsilon,1}$) of the experimental noise are given in Table \ref{tab:mean_hyper}. 
We also overlay 1000 simulations of Lindblad’s master equation on top of the experimental measurements, where each simulation corresponds to a sample from the Markov chains of Lindblad's parameters; see the top panels in Figure \ref{fig:ramsey01_all_cases} and Figure \ref{fig:ramsey12_all_cases}. 

\subsubsection{Approach 2: Bayesian Gaussian process characterization including model discrepancy}
\label{sec:ex2}

We next consider a discrepancy model, added externally to the simulation model as in \eqref{model_R0}-\eqref{model_R1}, and given by a Gaussian process with the exponential covariance kernel \eqref{cov_quantum} with $\gamma = 1$. 
We utilize the method of Section \ref{sec:marginal_likelihood}, retaining only the $r=50$ largest eigenvalues (and corresponding eigenvectors) of the covariance matrix, and marginalize over the remaining $n-r = 450$ eigenmodes. 

\begin{figure}[thb]
\centering
  \includegraphics[width=5.4cm]{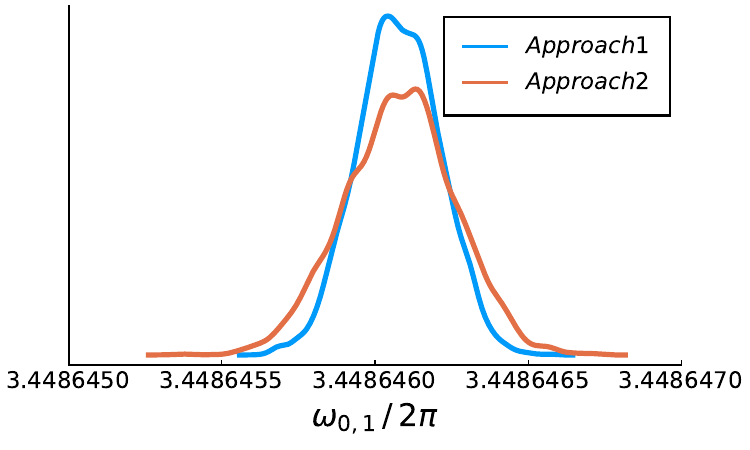}
  \includegraphics[width=5.4cm]{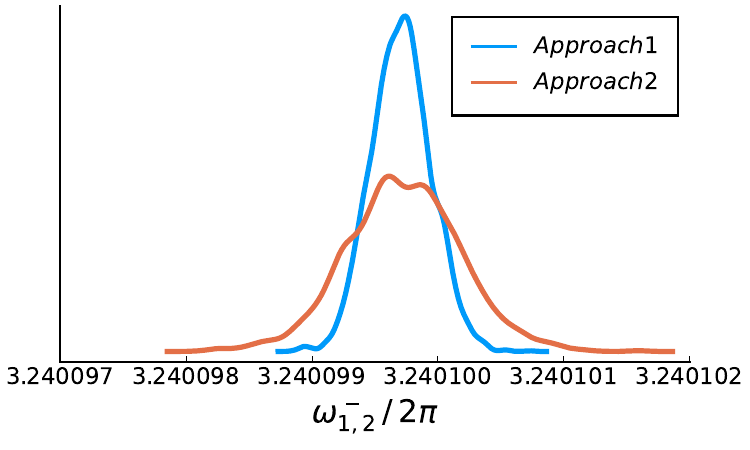}
  \includegraphics[width=5.4cm]{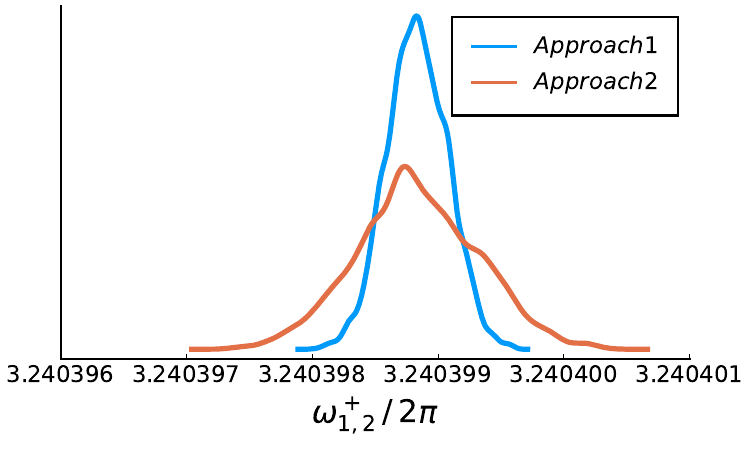}
  \\
  \vspace{0.9cm}
  \includegraphics[width=5.4cm]{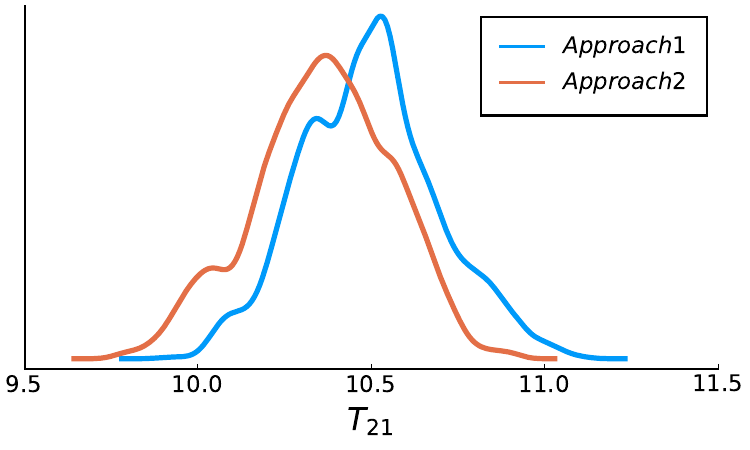} 
  \hskip 1cm
  \includegraphics[width=5.4cm]{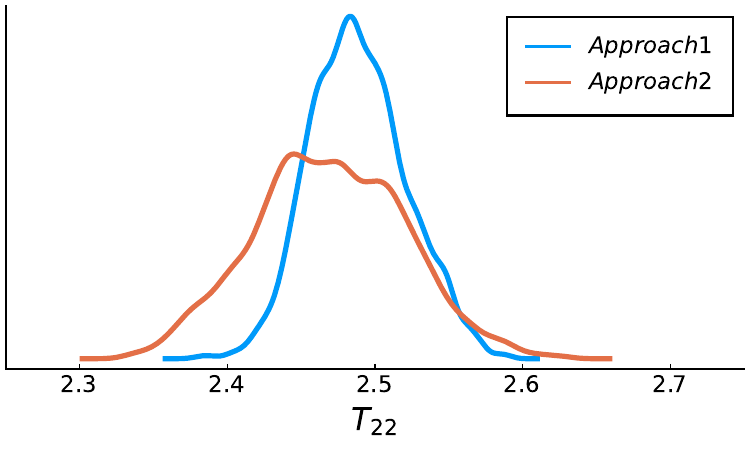}
\caption{Posterior densities of the five model parameters $\bt$ (transition frequencies and dephasing times) approximated by normalizing the Markov chain samples obtained by Approach \#1 (without model discrepancy) and Approach \#2 (with model discrepancy).} \label{fig:posteriors}
\end{figure}

\begin{figure}[thb]
\begin{center}
  \includegraphics[width=0.9\textwidth,height=4cm]{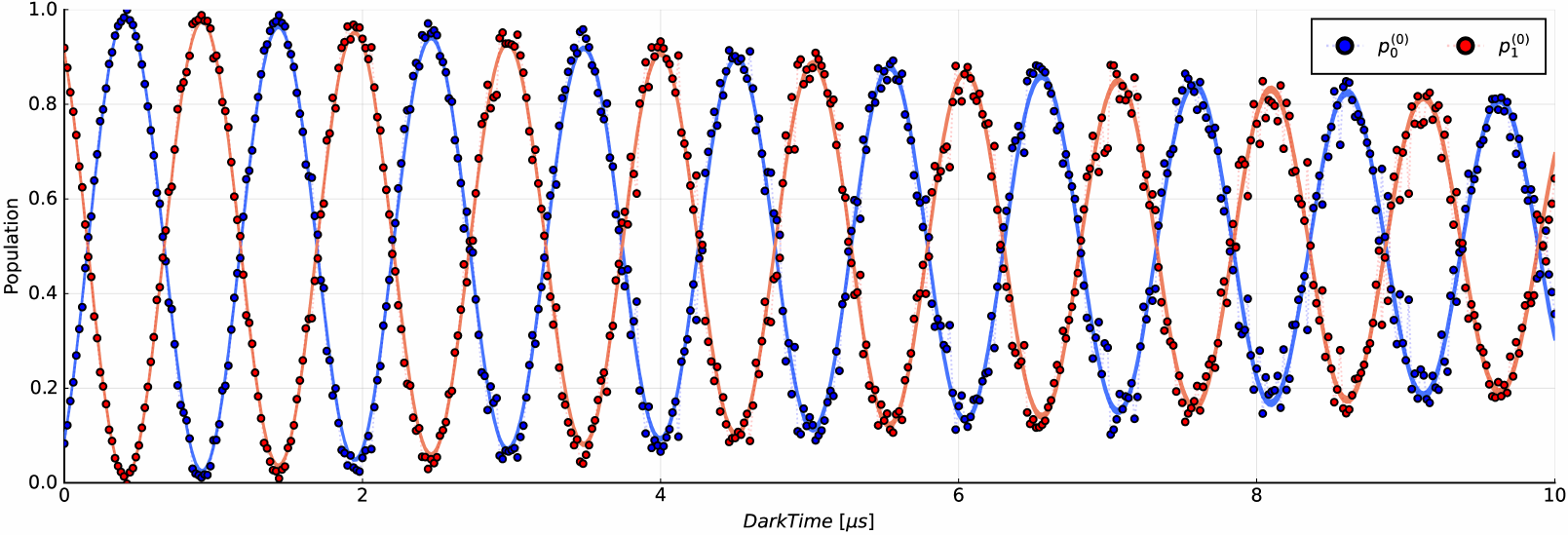}
  \includegraphics[width=0.9\textwidth,height=4cm]{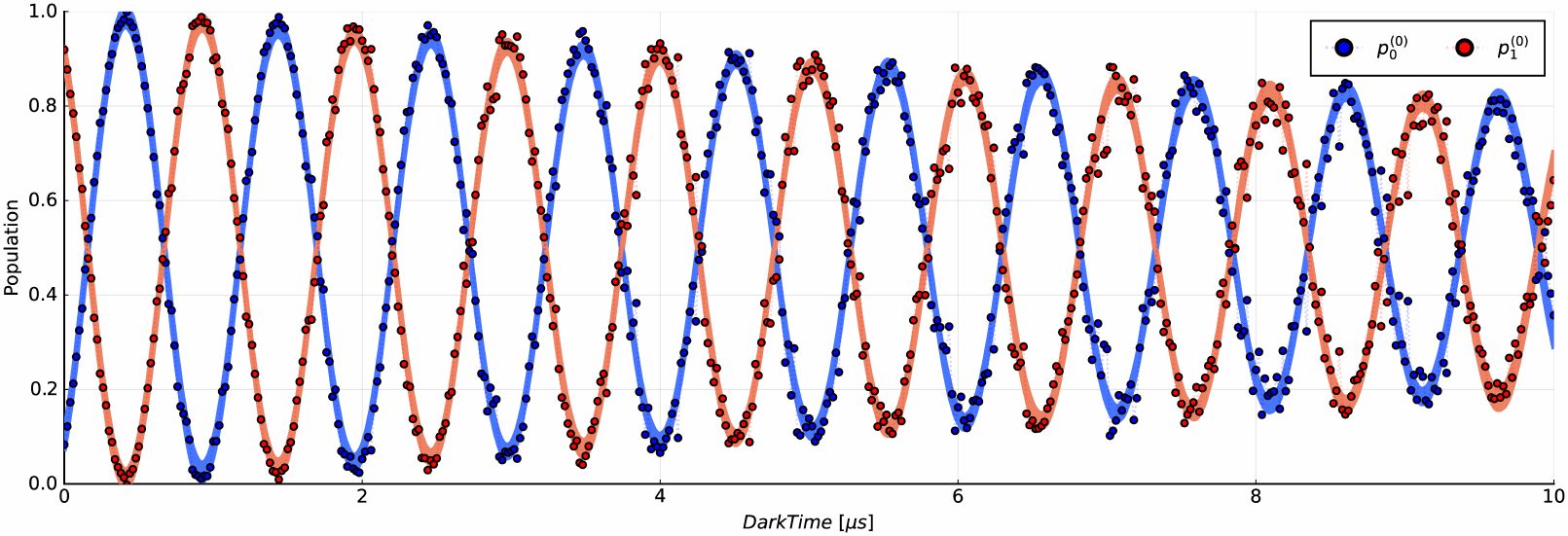}
\caption{Comparison between Ramsey $0\leftrightarrow 1$ experimental measurements (black circles) and 1000 realizations drawn from the predictive distributions (solid lines), obtained by Approach \#1 (top) and Approach \#2 (bottom). The variable thickness of the solid lines in the bottom panel is a result of including uncertainties from both measurement and model prediction in Approach \#2.} 
\label{fig:ramsey01_all_cases}
\end{center}
\end{figure}

\begin{figure}[thb]
\begin{center}
  \includegraphics[width=0.9\textwidth,height=4cm]{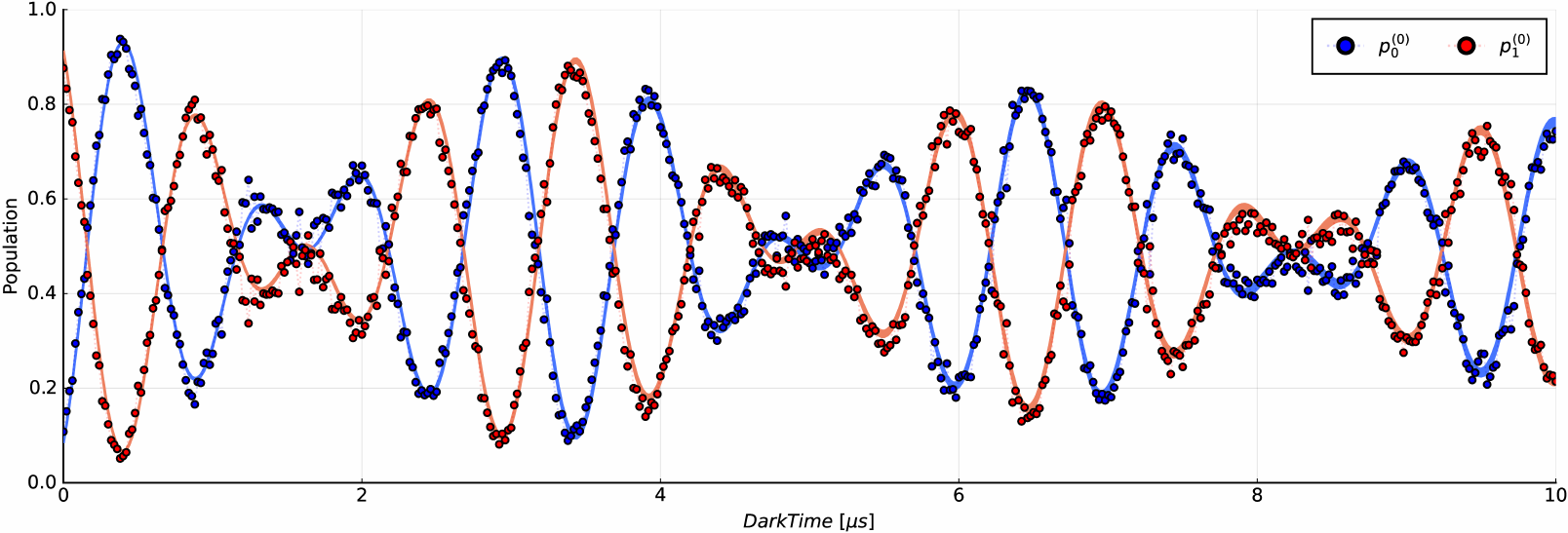}
  \includegraphics[width=0.9\textwidth,height=4cm]{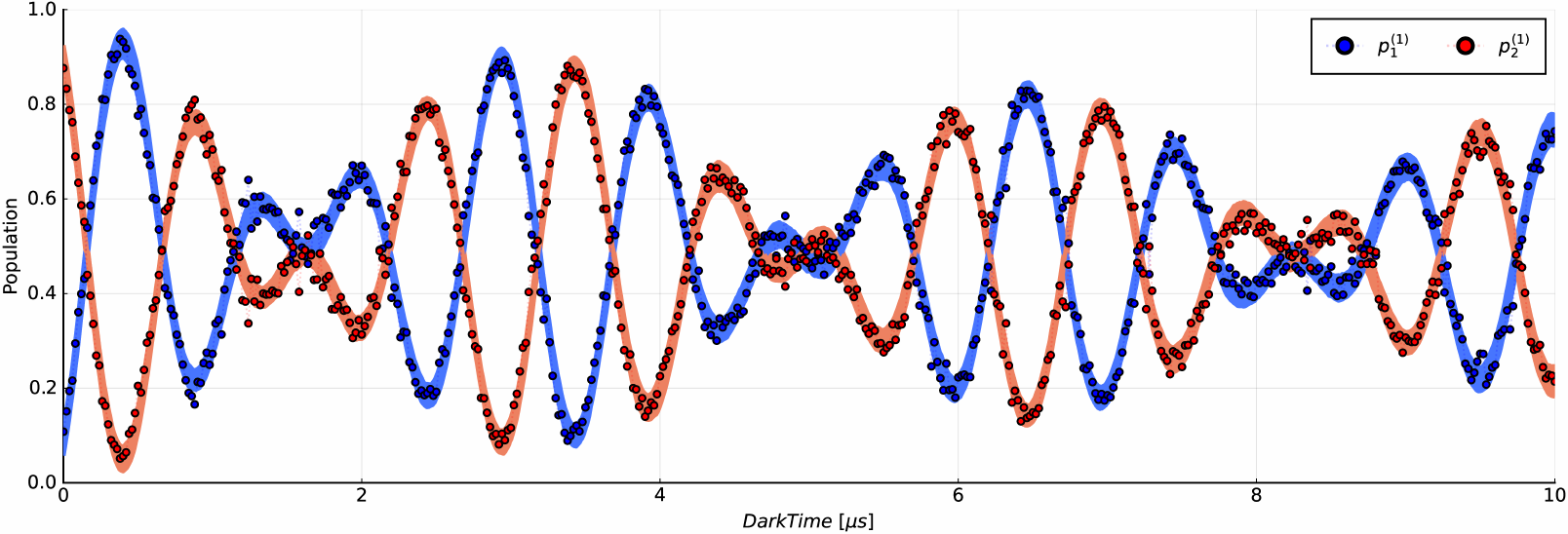}
\caption{Comparison between Ramsey $1\leftrightarrow 2$ experimental measurements (black circles) and 1000 realizations drawn from the predictive distributions (solid lines), obtained by Approach \#1 (top) and Approach \#2 (bottom). The variable thickness of the solid lines in the bottom panel is a result of including uncertainties from both measurement and model prediction in Approach \#2.}
\label{fig:ramsey12_all_cases}
\end{center}
\end{figure}

The posterior densities and the mean and standard deviation of the five model parameters $\bt$ can be found in Figure \ref{fig:posteriors} and Table \ref{tab:stat}, respectively. 
The mean of the hyper-parameters, i.e., the standard deviations of experimental noise and the covariance kernel parameters, are given in Table \ref{tab:mean_hyper}. 
We also overlay 1000 realizations of the predictive Gaussian process overlaid on top of the experimental measurements, see the bottom panels in Figure \ref{fig:ramsey01_all_cases} and Figure \ref{fig:ramsey12_all_cases}. 
Each realization is obtained by solving Lindblad's equation using a posterior sample of system parameters and adding a realization of the zero-mean Gaussian process with a deterministic covariance kernel obtained from the mean of the hyper-parameters, as explained in Section \ref{sec:prediction}.

%
\begin{table}[h!]
\begin{center}
  \begin{tabular}{|c|l|l|l|l|}
    \hline
    \multirow{2}{*}{Parameters} &
      \multicolumn{2}{c|}{Mean} &
      \multicolumn{2}{c|}{Standard Deviation} \\
    & no GP &  GP/w EK  & no GP & GP/w EK \\
    \hline
    $\omega_{01}/2\pi$  & $\SI{3.448646}{\giga\hertz}$  & $\SI{3.448646}{\giga\hertz}$  & $\SI{0.13}{\kilo\hertz}$ &  $\SI{0.19}{\kilo\hertz}$ \\
    \hline
    ${\omega}^-_{12} / 2\pi$ & $\SI{3.240100}{\giga\hertz}$  & $\SI{3.240100}{\giga\hertz}$ & $\SI{0.23}{\kilo\hertz}$ & $\SI{0.46}{\kilo\hertz}$ \\
    \hline
    ${\omega}^+_{12} / 2\pi$ & $\SI{3.240399}{\giga\hertz}$  & $\SI{3.240399}{\giga\hertz}$ & $\SI{0.23}{\kilo\hertz}$ & $\SI{0.48}{\kilo\hertz}$ \\
    \hline
    $T_{2,1}$ & 10.44 $\mu$s & 10.36 $\mu$s & 0.162 $\mu$s & 0.200 $\mu$s \\
    \hline
    $T_{2,2}$ & 2.49 $\mu$s & 2.47 $\mu$s & 0.027 $\mu$s & 0.051 $\mu$s \\
    \hline
  \end{tabular}
  \caption{Statistical summaries, including mean and standard deviation of transition frequencies and dephasing times determined by the two Bayesian characterization approaches. Here, GP stands for Gaussian process, and EK is the exponential covariance kernel.  \label{tab:stat}}
\end{center}
\end{table}

\begin{table}[h!]
\begin{center}
  \begin{tabular}{|c|c|c|}
    \hline
    \multirow{2}{*}{Hyper-parameters} &
      \multicolumn{2}{c|}{Mean} \\
    & no GP & GP/w EK \\
    \hline
    $\sigma_{\varepsilon,0}$  & 0.0387 & 0.0504 \\
    \hline
    $\sigma_{\delta,0}$  & --  & 0.0331 \\
    \hline
    $\tau_{0}$  & --  & 4.389 $\mu$s \\
    \hline
    \hline
    $\sigma_{\varepsilon,1}$  & 0.0307 & 0.0599 \\
    \hline
    $\sigma_{\delta,1}$  & -- & 0.0369 \\
    \hline
    $\tau_1$  & -- & 2.400 $\mu$s \\
    \hline
  \end{tabular}
  \caption{Mean of hyper-parameters by the two Bayesian characterization approaches: no Gaussian process (GP) and a GP with the exponential covariance kernel (EK).  \label{tab:mean_hyper}}
\end{center}
\end{table}



\subsection{Discussion}
\label{sec:discuss}

As Figure \ref{fig:posteriors} and Table \ref{tab:stat} show, the standard deviations of the system parameters obtained by Approach 2 are larger than those obtained by Approach 1. The larger standard deviation of Approach 2 is due to the spreading of noise in the measured data; the uncertainty is present in the data and is significant. Indeed, Approach 2 better captures this large uncertainty and characterizes it through larger standard deviations. Although the actual standard deviations are not known, one can conclude from Figures \ref{fig:ramsey01_all_cases}-\ref{fig:ramsey12_all_cases} that the predicted uncertainty model of Approach 2 better covers the spread of noise in the measurements compared to Approach 1. 

In Approach 1, we observe that while the predicted simulations follow the measurement trajectories on average, they fail to capture the uncertainty present in the measurements. 
In other words, although the uncertain model predictions are rather consistent in the mean with the measurements, their degree of uncertainty is inconsistent with the discrepancy from the measurements. 
This is also apparent from the very small noise variances (see the second column of Table \ref{tab:mean_hyper}). 
There may be two reasons for this failure. One reason is that a Gaussian likelihood model that is built upon the assumption that the noise is independently and identically distributed may not be capable of representing the complex noise structure of Ramsey measurements. 
A second reason, that is the subject of our focus here, is due to the inadequacy of Lindblad's model in describing the quantum system. 
As reported in \cite{Brynjarsdottir_OHagan:2014}, not accounting for model discrepancy may lead to ``overconfident" parameter estimates that in turn produce predictions with misleadingly small error ranges, a phenomenon that we indeed observe here.

In approach 2, the predicted simulations show better qualitative agreement and higher consistency with experimental measurements and the degree of uncertainty in the measurements, compared to the case where model discrepancy is ignored. 
This suggests that including model discrepancy is important for making reliable and consistent predictions.

We emphasize that GPs are useful statistical tools for capturing uncertainty. As a result, the aim of including a GP model discrepancy function in the present work is not to capture the missing physics or any missing terms in the dynamical model of the system. Here, the missing physics is considered as a type of (epistemic) uncertainty caused by the lack of information, and the GP model is added to account for the missing physics and to capture the uncertainty caused by it. This helps us quantify the level of uncertainty that is present in the quantum simulations due to the missing physics. 
We note that the GP model presented here does not fully capture uncertainty, particularly that of epistemic type. This is because epistemic uncertainty, by nature, lacks an inherent information structure, while a GP model imposes a random structure on the uncertainty. 
Moreover, the ability of GPs within the KOH framework to characterize more complex quantum systems and experimental scenarios depends on factors such as the number and connectivity of the qubits/qudits and the structure of SPAM errors. While our proposed framework is constrained by the inherent limitations of the KOH approach, the GP model, as demonstrated, improves predictive capability compared to the scenario where model uncertainty is completely disregarded.

\section{Conclusions}
\label{sec:conclusion}

We have introduced a stable and efficient marginal likelihood strategy within the KOH Bayesian framework, addressing computational challenges arising in the moderate-to-large data regime. By integrating out the degenerate eigenspace of the Gaussian process covariance matrix, our approach yields an exact likelihood on a lower-dimensional subspace, preserving all relevant inferential information while enhancing numerical stability and computational tractability. This framework is especially well-suited to applications in which the number of dominant eigenvalues is small relative to the problem dimension—a setting common in high-dimensional, strongly correlated datasets. As such, it offers a broadly applicable tool for uncertainty quantification and predictive modeling across domains where Gaussian processes are used, including Kriging, image analysis, and machine learning regression. Understanding the interplay between data distribution, eigenvalue decay, and model complexity remains an important direction for future research.

We demonstrated the utility of this framework through a Bayesian quantum characterization study of a tantalum-based transmon device in LLNL’s QuDIT system. Our results show that explicitly accounting for model-form discrepancy substantially improves predictive accuracy, enabling the quantum simulation model to reflect uncertainties beyond experimental noise. This is particularly important in quantum characterization, where uncertainty quantification is driven by model-form and experimental uncertainties rather than data scarcity. Our findings further underscore that ignoring model-form uncertainty can lead to overconfident predictions and underestimation of true system variability.

Despite these advances, several challenges remain for future research. A key limitation of the current framework is that model discrepancy is introduced externally via a Gaussian process, which can sometimes result in physically inconsistent predictions, such as populations falling below zero or exceeding one. Moreover, since the discrepancy term is learned from specific experimental data (e.g., Ramsey measurements), it may not generalize well to other experimental settings or observables. To address these limitations, future work will explore internal representations of model discrepancy, embedding it directly within the simulation model to enhance generalizability. Such an approach could enable more robust quantum optimal control strategies, leading to control pulses with improved resilience to noise.

Another critical challenge in quantum characterization is scalability. The density matrix dimension scales exponentially with the number of qubits, while system parameters grow polynomially, making traditional Bayesian inference methods computationally prohibitive. Addressing this requires developing surrogate models for Lindblad’s equation and scalable alternatives to MCMC sampling, such as variational inference techniques. These advancements will be crucial for extending our framework to larger quantum systems, further bridging the gap between verification, validation, and uncertainty quantification in quantum computing applications.

\vspace{-0.1cm}
\section*{Acknowledgements}
The first author was supported by the U.S. Department of Energy, Office of Science, Advanced Scientific Computing Research, under Award No. DE-SC0025481, and by the National Science Foundation under Grant No. DMS-2436318. 
Both authors gratefully acknowledge support from the U.S. Department of Energy, Office of Science, Advanced Scientific Computing Research, through the Advanced Research in Quantum Computing program, project TEAM, award SCW-1683.1. 
The authors also thank Dr.Gabriel Huerta (Sandia National Laboratories) for his guidance and stimulating discussions, and Dr. Yujin Cho (Lawrence Livermore National Laboratory) for her insights into the experimental intricacies of the QuDIT device. 
This work was performed under the auspices of the U.S. Department of Energy by Lawrence Livermore National Laboratory under Contract DE-AC52-07NA27344. This is contribution LLNL-JRNL-2006155.\footnote{This manuscript has been authored by Lawrence Livermore National Security, LLC under Contract No.~DE-AC52-07NA27344 with the US.~Department of Energy. The United States Government retains, and the publisher, by accepting the article for publication, acknowledges that the United States Government retains a non-exclusive, paid-up, irrevocable, world-wide license to publish or reproduce the published form of this manuscript, or allow others to do so, for United States Government purposes.}

\vspace{-0.1cm}


\bibliography{refs_motamed}

\bibliographystyle{unsrt}

\end{document}